\documentclass[11pt]{article}
\usepackage{graphicx}
\usepackage{amssymb}
\usepackage{amsmath}
\usepackage{graphicx}
\usepackage{amstext}
\usepackage{esint}
\usepackage[utf8]{inputenc}
\usepackage{color}
\usepackage{cite}

\numberwithin{equation}{section}

\def\beq{\begin{eqnarray}}    
\def\eeq{\end{eqnarray}}      

\newcommand{\rL}{\rho_\Lambda}

\newcommand{\CC}{\Lambda}

\newcommand{\rv}{\rho_{\rm vac}}
\newcommand{\rvo}{\rho^0_{\rm vac}}

\newcommand{\OLo}{\Omega^0_{\Lambda}}

\newcommand{\rco}{\rho^0_{c}}

\newcommand{\nueff}{\nu_{\rm eff}}

\newcommand{\bk}{{\bf k}}

\newcommand{\cH}{\mathcal{H}}
\newcommand{\cpH}{\mathcal{H}^\prime}

\newcommand{\joantext}[1]{{\textcolor{black}{#1}}}

\begin{document}



 \hyphenation{nu-cleo-syn-the-sis u-sing si-mu-la-te ma-king
cos-mo-lo-gy know-led-ge e-vi-den-ce stu-dies be-ha-vi-or
res-pec-ti-ve-ly appro-xi-ma-te-ly gra-vi-ty sca-ling
ge-ne-ra-li-zed re-gu-la-ri-za-tion mo-del mo-dels po-wers ex-cee-din-gly ho-we-ver}




\begin{center}
{\bf \Large Running vacuum in quantum field theory in curved spacetime:  renormalizing $\rho_{vac}$ without  $\sim m^4$ terms} \vskip 2mm

 \vskip 8mm

\textbf{Cristian Moreno-Pulido and  Joan Sol\`a Peracaula}

\vskip 0.5cm
Departament de F\'isica Qu\`antica i Astrof\'isica, \\
and \\ Institute of Cosmos Sciences,\\ Universitat de Barcelona, \\
Av. Diagonal 647, E-08028 Barcelona, Catalonia, Spain

\vskip0.5cm

\vskip0.4cm

E-mails:  cristian.moreno@fqa.ub.edu, sola@fqa.ub.edu

 \vskip2mm

\end{center}
\vskip 15mm

\begin{quotation}
\noindent {\large\it \underline{Abstract}}.
The $\Lambda$-term in Einstein's equations is  a fundamental building block of the `concordance'   $\Lambda$CDM
model of cosmology.  Even though the model is not free of fundamental problems,  they have not been circumvented by any  alternative dark energy proposal  either.  Here we stick to the $\Lambda$-term, but we contend that it can be a `running quantity' in quantum field theory (QFT) in curved spacetime.  A plethora of phenomenological works have shown that this option can be highly competitive with the  $\Lambda$CDM with a rigid cosmological term. The, so-called,  `running vacuum models' (RVM's) are characterized by the vacuum energy density,  $\rho_{vac}$,  being a  series of (even) powers of the Hubble parameter and its time derivatives.    Such theoretical form has been motivated by general  renormalization group arguments, which look plausible.  Here we  dwell further upon the origin of the RVM structure within QFT in FLRW spacetime. We compute the renormalized  energy-momentum tensor with the help of  the  adiabatic regularization procedure and find that it leads essentially to the  RVM form. This means that  $\rho_{vac}(H)$ evolves as a constant term plus  dynamical components ${\cal O}(H^2)$  and ${\cal O}(H^4)$, the latter being relevant for the early universe only.  However, the renormalized  $\rho_{vac}(H)$  does not  carry dangerous terms proportional to the quartic power of the masses ($\sim m^4$) of the fields, these terms being a well-known source of exceedingly large contributions.  At present, $\rho_{vac}(H)$  is dominated by the additive constant term accompanied by a mild dynamical component  $\sim \nu H^2$  ($|\nu|\ll1$), which mimics quintessence.
\end{quotation}
\vskip 5mm

\newpage

\tableofcontents

\newpage


\section{Introduction}\label{intro}

The cosmological constant (CC) term,  $\Lambda$,  in Einstein's equations has been for some three decades a fundamental building block of the `concordance' or standard  $\Lambda$CDM
model of cosmology\,\cite{Peebles1984}. The model, however, was phenomenologically favored only  as of the time $\CC$  became a physically measured quantity some twenty years ago\,\cite{SNIa}. Nowadays $\Lambda$, or more precisely the associated current cosmological parameter \joantext{$\OLo=\rL/\rco$} became a precision quantity\,\cite{PlanckCollab}. Here \joantext {$\rL=\CC/(8\pi G_N)$}  is the (vacuum) energy density induced by $\CC$, $G_N$ is Newton's constant and $\rco=3H_0^2/(8\pi G_N)$ is the current critical density.  The accurate knowledge of $\OLo$ around $0.7$ is an important observational achievement, but it does not mean that we fully understand its nature and origin at a fundamental level.  The cosmological constant problem\,\cite{Weinberg89,Witten2000} is a preeminent example of a fundamental theoretical conundrum, which actually affects all forms of dark energy (DE)\,\cite{Sahni2000,PeeblesRatra2003,Padmanabhan2003,Copeland2006,DEBook}.  The abstruse theoretical problems, though, are not the only nagging ones afflicting the concordance model. In practice the  $\Lambda$CDM appears to be currently in tension with some important measurements, most significantly the discordant values of the current Hubble parameter $H_0$ obtained independently from measurements of the local and the early universe\,\cite{TensionsLCDM}.  Whether these tensions are the result of as yet unknown systematic errors is not known, but there remains perfectly upright the possibility that a deviation from the $\CC$CDM model could provide an explanation for such discrepancies\cite{Riess2019}. As it has been shown in the literature,  models mimicking a time-evolving  $\CC$ (and hence a dynamical vacuum energy density $\rL$) could help in alleviating these problems, see e.g.\,\cite{ApJL2015,RVMphenoOlder1,RVMpheno1,RVMpheno2,ApJL2019,Mehdi2019,BD2020} and  \cite{Vagnozzi,Valentino,OobaRatra,ParkRatra,MartinelliSalvatelli,Costa,LiZhang}.

In this paper, we would like to further dwell upon the theoretical possibility of having a dynamical vacuum energy density (VED),  $\rv$,  in the framework of quantum field theory (QFT) in curved spacetime\,\cite{BirrellDavies82,ParkerToms09,Fulling89,MukhanovWinitzki07}.   Above all we wish to focus on the dynamics associated to the running vacuum model (RVM)\,\cite{ShapSol1,Fossil2008,ShapSol2}; for a review, see \cite{JSPRev2013,JSPRev2015,AdriaPhD2017} and references therein.  For related studies, see e.g. \cite{Babic2005,Maggiore2011} and  \cite{KohriMatsui2017,Antipin2017,Ferreiro2019,Ferreiro2020}, some of them extending the subject to the context of supersymmetric theories\,\cite{Bilic2011,Bilic2012} and also to supergravity\,\cite{SUGRA2015}.  More recently the matter has also been addressed successfully in the framework of the effective action of string theories\,\cite{BMS2020A,BMS2020B}. Here, however, we aim at the  computation of the VED in QFT in a curved background, specifically in the spatially flat Friedmann-Lema\^\i tre-Robertson-Walker (FLRW) metric. We proceed by renormalizing the energy-momentum tensor using the adiabatic regularization prescription (ARP)\,\cite{BirrellDavies82,ParkerToms09}. This renormalization method is based on  the WKB approximation of the field modes  in the expanding universe.  We perform the calculation in two ways, one through a modified form of the ARP\,\cite{Ferreiro2019} and the second  (presented in one appendix) involving dimensional regularization (DR). The common result is that the properly renormalized VED, obtained upon inclusion of the renormalized value of $\rL$ at a given scale, does not contain the unwanted contributions proportional to the fourth power of the particle masses ($\sim m^4$) and hence it is free from  large induced corrections to the VED. This is tantamount to subtracting the Minkowskian contribution from the curved spacetime result, as we show. In addition, we find that the final expression for  the  VED adopts the RVM form for the current universe, namely it contains not only the usual constant term but also one  that evolves with the square of the Hubble rate ($\sim \nu H^2$, with $|\nu|\ll 1$).  The latter represents only a mild (dynamical) correction to the constant contribution and it can mimic quintessence or phantom DE depending on the sign of $\nu$.

The structure of the article is as follows. In Sec. \ref{sec:EMT} we define our framework, which consists of a neutral scalar field non-minimally coupled to gravity, and compute the classical energy-momentum tensor (EMT). In Sections \ref{sec:AdiabaticVacuum} and \ref{sec:AREMT}  we address the quantum fluctuations in the adiabatic vacuum through the WKB expansion of the field modes in the  FLRW background. We discuss the adiabatic regularization of the EMT.
In Sec. \ref{sec:RenormZPE} we proceed to renormalize the EMT in the FLRW context using the adiabatic prescription, which is then needed  in  Sec. \ref{sec:RenormalizedVED} to extract the precise form of  $\rv$  from the renormalized zero-point energy (ZPE) up to terms of adiabatic order $4$,  which in our case means up to  ${\cal O}(H^4)$. We show that the relation between values of the VED at different scales is free from quartic powers of the masses. We also demonstrate that our renormalization procedure gives the same result as subtracting the Minkowskian contribution from the curved spacetime result. In Sec. \ref{sec:RunningConnection}, we provide the connection of the computed VED in this work with the running vacuum model (RVM), which had been derived before from the general point of view of the renormalization group in curved spacetime.  The final discussion and a summary of the conclusions is presented in Sec. \ref{sec:conclusions}.  Three appendices at the end furnish complementary material. Specifically, Appendix \ref{AppendixA} defines our conventions and collects some useful formulas.  Appendix \ref{AppendixB} reconsiders the main parts of the renormalization of the EMT using dimensional regularization and the standard counterterm procedure, starting of course from the same WKB expansion of the field modes. Finally, Appendix \ref{AppendixC} discusses alternative identifications of the VED leading to generalized forms of the RVM which had already been anticipated from the renormalization group approach in previous works.

\section{Energy-Momentum tensor for non-minimally coupled scalar field}\label{sec:EMT}

The gravitational field equations read\,\footnote{A list of geometric quantities of interest here are shown in the Appendix \ref{AppendixA}, where we also define our conventions.}
\begin{equation}
R_{\mu \nu}-\frac{1}{2}R g_{\mu \nu}+\Lambda g_{\mu \nu}=8\pi G_N T_{\mu \nu}^{\rm matter}\,, \label{FieldEq}
\end{equation}
where $ T_{\mu \nu}^{\rm matter}$  is the EMT of matter.
They can conveniently be rewritten as
\begin{equation} \label{FieldEq2}
\frac{1}{8\pi G_N}G_{\mu \nu}+\rho_\Lambda g_{\mu \nu}=T_{\mu \nu}^{\rm matter}\,,
\end{equation}
where  $\rho_\Lambda \equiv \Lambda/(8\pi G_N)$ is the VED associated to $\CC$. The latter contributes a term $T_{\mu \nu}^\Lambda \equiv -\rho_\Lambda g_{\mu \nu}$  to the total EMT.  However, in general, there will be also other contributions to the total VED, in particular those associated to the quantum fluctuations of the fields, and also to their  classical ground state energy (if it is nonvanishing).  For simplicity we will suppose that there is only one (matter) field contribution to the EMT  on the right hand side of \eqref{FieldEq2} in the form of  a real scalar field, $\phi$, and such contribution will be denoted $T_{\mu \nu}^{\phi}$.  Hence the total EMT reads  $T^{\rm tot}_{\mu\nu}=T_{\mu \nu}^\Lambda+T_{\mu \nu}^{\phi}$. We neglect the incoherent matter contributions (e.g. from dust and radiation) for the kind of QFT considerations made in this study, as they can be added without altering the QFT aspects.

Suppose that the scalar field is non-minimally coupled to gravity and that it does not couple to itself\,\footnote{We will not consider  a possible contribution from a classical potential for $\phi$ in our analysis, which in general should also involve quantum corrections and hence leading to an effective potential.  Here we wish to concentrate mainly on the zero-point energy of the quantum fields, which in itself is already rather cumbersome.}. The part of the action involving $\phi$,  then, reads
\begin{equation}\label{eq:Sphi}
  S[\phi]=-\int d^4x \sqrt{-g}\left(\frac{1}{2}g^{\mu \nu}\partial_{\nu} \phi \partial_{\mu} \phi+\frac{1}{2}(m^2+\xi R)\phi^2 \right)\,,
\end{equation}
where $\xi$ is the non-minimal coupling of $\phi$ to gravity.  In the special case $\xi=1/6$, the massless ($m=0$)  action is conformally invariant, i.e. symmetric under simultaneous local Weyl rescalings of the metric and the scalar field: $g_{\mu\nu}\to e^{2\alpha(x)}g_{\mu\nu}$ and  $\phi\to e^{-\alpha(x)}\phi$, for any local spacetime function $\alpha(x)$.
However, we will keep $\xi$ general since our scalar field will be massive.

In general,  the non-minimal coupling $\xi$ is needed for renormalization since it is generated by loop effects even if it is absent in the classical action\,\cite{BirrellDavies82}. However, $\xi$ is not needed for the renormalization of the action in the present case since the scalar field is free as a quantum field, its interaction being only with the classical geometric/gravitational background --  see the aforementioned footnote on this page.  More details are put forward in Appendix\,\ref{AppendixB}, where an explicit counterterm renormalization procedure is employed.  However, by keeping $\xi\neq 0$ we can provide more general results, which will be particularly useful for the connection of our calculations with the RVM framework in Sec.\ref{sec:RunningConnection}. In addition, the presence of a non-minimal coupling is expected in a variety of contexts of extended gravity theories\,\cite{Sotiriou2010,Capozziello2011,CapozzielloFaraoni2011}. For instance,  $f(R)$ gravity is equivalent to scalar-tensor theory,  and also to Einstein theory coupled to an ideal fluid\,\cite{Capozziello2006}.  The non-minimal coupling is crucially involved in models of Higgs-induced inflation\,\cite{Barvinsky2008}. Furthermore,  higher order and non-minimally coupled terms can be transformed, by means of a conformal transformation, into Einstein gravity plus one or more scalar fields minimally coupled to curvature.  These are only a few examples in QFT,  see e.g.\,\cite{CapozzielloFaraoni2011} and references therein. Let us also mention that non-minimal coupling of dilaton fields to curvature  are also common  in the context of the effective action of string theory at low energies --  see  Sec.\,\ref{sec:RVMinflation} for an interesting connection of the RVM with strings.

The field $\phi$ obeys the  Klein-Gordon (KG) equation
\begin{equation}
(\Box-m^2-\xi R)\phi=0\,,\label{KGequation}
\end{equation}
 where $\Box\phi=g^{\mu\nu}\nabla_\mu\nabla_\nu\phi=(-g)^{-1/2}\partial_\mu\left(\sqrt{-g}\, g^{\mu\nu}\partial_\nu\phi\right)$.
 In the case of general non-minimal coupling $\xi$, the EMT can be computed upon straightforward calculation:
\begin{equation}
\begin{split}
T_{\mu \nu}(\phi)=&-\frac{2}{\sqrt{-g}}\frac{\delta S_\phi}{\delta g^{\mu\nu}}= (1-2\xi) \partial_\mu \phi \partial_\nu\phi+\left(2\xi-\frac{1}{2} \right)g_{\mu \nu}\partial^\sigma \phi \partial_\sigma\phi\\
& -2\xi \phi \nabla_\mu \nabla_\nu \phi+2\xi g_{\mu \nu }\phi \Box \phi +\xi G_{\mu \nu}\phi^2-\frac{1}{2}m^2 g_{\mu \nu} \phi^2.
\end{split} \label{EMTScalarField}
\end{equation}
In the following, we are going to consider the spatially flat FLRW metric in the conformal frame. Introducing the conformal time, $\eta$, we have  $ds^2=a^2(\eta)\eta_{\mu\nu}dx^\mu dx^\nu$, with $\eta_{\mu\nu}={\rm diag} (-1, +1, +1, +1)$ the Minkowski  metric.
We will denote the derivative with respect to the conformal time by  $^\prime\equiv d/d\eta$. The corresponding Hubble rate  is  $\mathcal{H}(\eta)\equiv a^\prime /a$.  Since $dt=a d\eta$, the usual Hubble rate with respect to the cosmic time,   $H(t)=\dot{a}/a$ (with $\dot{}\equiv d/dt$),  is related to the former through  $\mathcal{H}(\eta)=a  H(t)$. We will present most of our calculations in terms of the conformal time, but at the end it will be useful to express the VED in terms of the usual Hubble rate $H(t)$, as this will ease the comparison with the RVM results in the literature.

Because our metric is conformally flat,  $g_{\mu\nu}=a^2(\eta)\eta_{\mu\nu}$, we have the inverse $g^{\mu\nu}=a^{-2}(\eta)\eta^{\mu\nu}$ and $\sqrt{-g}=a^4(\eta)$, and as a result the action \eqref{eq:Sphi} can be rewritten
\begin{equation}\label{eq:Sphi2}
  S[\phi]=\frac12\int d\eta\,d^3x\, a^2 \left(\phi^{\prime 2} -(\nabla\phi)^2  - a^2(m^2+\xi R)\phi^2 \right)\,.
\end{equation}
If we perform the field redefinition $\phi=\varphi/a$ and disregard total derivatives, the previous action becomes the following functional of $\varphi$:
\begin{equation}\label{eq:Svarphi}
  S[\varphi] =\frac12\int d\eta d^3x \left\{\varphi^{\prime 2} -(\nabla\varphi)^2  - a^2 \left[m^2 + \left(\xi-\frac16\right) R \right] \varphi^2\right\}\,,
\end{equation}
where we have used (cf. Appendix \ref{AppendixA}) $R=6a^{\prime\prime}/a^3$.  The above field redefinition enables us to have a simpler field equation  for $\varphi$ as if we were in Minkowski space (with conformal time) and an effective time-dependent mass different from that in  \eqref{KGequation}.  Computing $\delta S[\varphi]/\delta\varphi=0$ from  (\ref{eq:Svarphi}) we find:
\begin{equation}
(\tilde{\Box}-m^2_{\rm eff}(\eta))\varphi=0\,,\ \ \ \ \ \ \  m^2_{\rm eff}(\eta)\equiv  a^2(\eta) \left[m^2 + \left(\xi-\frac16\right) R(\eta) \right]\,,\label{KGvarphi}
\end{equation}
 where $\tilde{\Box}\varphi \equiv \eta^{\mu\nu}\partial_\mu\partial_\nu\varphi=-\varphi''+\nabla^2\varphi$ is the Minkowskian box operator acting on $\varphi$ in conformal coordinates $x^\mu=(\eta, {\bf x})$. The above equation for $\varphi$  is, of course, equivalent to \eqref{KGequation} for  the original field $\phi$, as one can check by computing the curved spacetime box operator in the conformal metric.

 \section{Quantum fluctuations and adiabatic vacuum}\label{sec:AdiabaticVacuum}

 Let us now move from classical to quantum field theory. We can take into account the quantum fluctuations of the field $\phi$  by considering the expansion of the field around its background (or classical mean field) value $\phi_b$:
\begin{equation}
\phi(\eta,x)=\phi_b(\eta)+\delta \phi (\eta,x). \label{ExpansionField}
\end{equation}
We identify the vacuum expectation value (VEV) of the field with the background value,  that is to say, $\langle 0 | \phi (\eta, x) | 0\rangle=\phi_b (\eta)$, whereas we assume zero VEV for the fluctuation:  $\langle 0 | \delta\phi | 0\rangle =0$ (as it will become evident from the mode expansion in terms of creation and annihilation operators to be discussed below). We will define the vacuum state to which we are referring to with more precision below.

Given the above field decomposition into a classical plus fluctuating part, the corresponding EMT decomposes itself as $\langle T_{\mu \nu}^\phi \rangle=\langle T_{\mu \nu}^{\phi_b} \rangle+\langle T_{\mu \nu}^{\delta_\phi}\rangle$, where
$\langle T_{\mu \nu}^{\phi_{b}} \rangle =T_{\mu \nu}^{\phi_{b}} $
is the  contribution  from the classical or background part,  whereas  $\langle T_{\mu \nu}^{\delta_\phi}\rangle$  is  the contribution from the quantum fluctuations. The $00$-component of the latter is connected with the zero-point energy (ZPE) density of the scalar field in the FLRW background. Thus, the total vacuum contribution to the EMT reads
\begin{equation}
\langle T_{\mu \nu}^{\rm vac} \rangle= T_{\mu \nu}^\Lambda+\langle T_{\mu \nu}^{\delta \phi}\rangle=-\rho_\Lambda g_{\mu \nu}+\langle T_{\mu \nu}^{\delta \phi}\rangle\,.\label{EMTvacuum}
\end{equation}
The above equation says that the total vacuum EMT  is made out of the contributions from the cosmological term and of the quantum fluctuations  of the field.  We will use later on a renormalized version of this equation and extract a relation satisfied by the renormalized VED. \joantext{Notice that we use the notation $\rho_\Lambda=\Lambda/(8\pi G_N)$  to denote a parameter in the Einstein-Hilbert action.  This is not yet the physical vacuum energy density, $\rho_{\rm vac}$, which we are aiming at. The latter is obtained from the $00$-component of the l.h.s. of Eq. (3.2) --  see Sec.\,\ref{sec:RenormalizedVED}  for its precise definition and Appendix C for an extended discussion.  In this respect, let us note that in the introduction we have denoted the physical quantity in the conventional form $\rho_\Lambda$, but this should not be confused with the more precise notations used hereafter.}

The field (\ref{ExpansionField}) obeys the  curved spacetime KG equation (\ref{KGequation}) independently  by the classical and  quantum parts.  Similarly,  $\varphi$ and $\delta\varphi$ obey separately the Minkowskian KG equation (\ref{KGvarphi}).   Let us concentrate on the fluctuation $\delta\varphi$.  We can decompose it in Fourier frequency modes $h_k(\eta)$:
\begin{equation}
\delta \varphi(\eta,{\bf x})=\frac{1}{(2\pi)^{3/2}}\int d^3{k} \left[ A_\bk e^{i{\bf k\cdot x}} h_k(\eta)+A_\bk^\dagger e^{-i{\bf k\cdot x}} h_k^*(\eta) \right]\,. \label{FourierModes}
\end{equation}
Since  $\phi=\varphi/a$,  the expansion of $\delta\phi$ is, of course, the same as that of (\ref{FourierModes}) but divided by the scale factor $a$.
Here  $A_\bk$ and  $A_\bk^\dagger $ are the (time-independent) annihilation and creation operators.
Their commutation relations are the usual ones
\begin{equation}
[A_\bk, A_\bk'^\dagger]=\delta({\bf k}-{\bf k'}), \qquad [A_\bk,A_ \bk']=0. \label{CommutationRelation}
\end{equation}
Notice that  $A_\bk$ and  $h_k$  have mass dimensions $-3/2$ and $-1/2$  in natural units, respectively.
Upon substituting  the Fourier expansion (\ref{FourierModes})  in  $(\tilde{\Box}-m^2_{\rm eff}(\eta))\delta\varphi=0$ we find that the frequency modes of the fluctuations satisfy the (linear) differential equation
\begin{equation}
h_k^{\prime \prime}+ \Omega_k^2 h_k=0\,, \ \ \ \ \ \ \ \ \ \ \Omega_k^2(\eta) \equiv k^2+m^2_{\rm eff}(\eta)=\omega_k^2(m)+a^2\, (\xi-1/6)R\,, \label{KGModes}
\end{equation}
with  $\omega_k^2(m)\equiv k^2+a^2 m^2$.
As we can see, $h_k$ depends only on the modulus $k\equiv|\bk|$ of the momentum.
Because $\Omega_k(\eta)$ is a nontrivial function of the conformal time, the modes cannot be found in a simple form. However, one can generate an approximate solution from  a recursive self-consistent iteration based on the phase integral ansatz
\begin{equation}
h_k(\eta)=\frac{1}{\sqrt{2W_k(\eta)}}\exp\left(i\int^\eta W_k(\tilde{\eta})d\tilde{\eta} \right)\,. \label{WKBSolution}
\end{equation}
The latter is normalized through the Wronskian condition
$ h_k^\prime h_k^* -  h_k^{} h_k^{*\prime}=i$, 
which insures that the standard equal-time commutation relations between the field operator $\varphi$ and its canonical momentum, $\pi_\varphi=\varphi^\prime$, are preserved.
The function $W_k$ in the above ansatz is solution of the differential equation obtained from inserting \eqref{WKBSolution} into \eqref{KGModes}:
\begin{equation}
W_k^2=\Omega_k^2 -\frac{1}{2}\frac{W_k^{\prime \prime}}{W_k}+\frac{3}{4}\left( \frac{W_k^\prime}{W_k}\right)^2\,. \label{WKBIteration}
\end{equation}
Although this equation is non-linear, it can be solved using the WKB approximation. Taking into account that the WKB solution is  valid for large $k$ (i.e. for short wave lengths, as e.g. in geometrical Optics) the function  $\Omega_k$ is slowly varying for weak fields. This motivates  a notion of vacuum  called the adiabatic vacuum\,\cite{Bunch1980}. Rather than formulating it as the state without particles, we can at least say it is a state essentially empty of high frequency modes.  Indeed, particles with definite frequencies cannot be strictly defined in a curved background, since $\Omega_k(\eta)$ is a function of time.  Nonetheless a Fock space interpretation is still possible, and the adiabatic vacuum can be formally defined as the quantum state which is annihilated by all the operators $A_k$ of the above Fourier expansion, see\,\cite{BirrellDavies82,ParkerToms09,Fulling89,MukhanovWinitzki07} for details.  Our VEV's actually refer to that adiabatic vacuum.  In such conditions, the minimal excited state is $h_k\simeq e^{ik\eta}/\sqrt{2k}$, with $k\simeq\Omega_k$, and hence  one can maintain an approximate particle interpretation of the quantized fields in a curved background provided the geometry  is slowly varying.  However, in general,  the physical interpretation of the modes (\ref{KGModes}) with time varying frequencies must be sought in terms of field observables rather than in particle language.  In practice, the adiabatic vacuum approximation assumes both short wavelengths and weak (or at least non strong) gravitational fields, such  that the effective frequencies $\Omega_k$ are slowly varying functions of  time around the Minkowskian values defined through the masses and momenta. Therefore both $m_{eff}^2$ and   $\Omega_k^2$ remain safely positive in our domain of study. Simple estimates show that this is so for the most accessible part of the cosmic history, starting from the radiation-dominated epoch (where $R=0$) until the present time and into the future,  in which $R\sim H^2$ is very small as compared to the usual particle masses (squared).  We emphasize that in all cases, including  the situation with the stronger gravitational fields in the  inflationary epoch (see Sec.\,\ref{sec:RVMinflation} for further discussion),  a more physical interpretation of the vacuum effects of the expanding universe can be achieved by computing the  renormalized EMT in the FLRW background. The first thing to do in order to carry out this task in an efficient way in our case is to (adiabatically) regularize the EMT.

\section{Adiabatic Regularization of the Energy-Momentum tensor}\label{sec:AREMT}

For an adiabatic (slowly varying) $\Omega_k$, we can use Eq.\,(\ref{WKBIteration}) as a recurrence relation to generate an (asymptotic) series solution. In the gravitational context, such  WKB approximation  is organized through adiabatic orders and constitutes the basis for the  adiabatic regularization procedure (ARP)\footnote{The ARP was first introduced for minimally coupled (massive) scalar fields in\,\cite{ParkerFulling1974,BunchParker1979} and subsequently generalized for arbitrary couplings\,\cite{Bunch1980}.
 For a review, see e.g.  the classic books \cite{BirrellDavies82,ParkerToms09}. The method has been applied to related studies of QFT in curved backgrounds\,\cite{Ferreiro2019,KohriMatsui2017} and has also been extended for spin one-half fields in\,\cite{Landete,Barbero}.}. The quantities that are taken to be of adiabatic order 0 are: $k^2$ and $a$. Of adiabatic order 1 are: $a^\prime$ and $\mathcal{H}$. Of adiabatic order 2: $a^{\prime \prime},a^{\prime 2},\mathcal{H}^\prime$ and $\mathcal{H}^2$. Each additional derivative increases the adiabatic order  by one unit.
Therefore, the solution of the ``effective frequency'' $W_k$  is found from a WKB-type asymptotic expansion in powers of the adiabatic order:
\begin{equation}\label{WKB}
W_k=\omega_k^{(0)}+\omega_k^{(2)}+\omega_k^{(4)}\dots,
\end{equation}
where each $\omega_k^{(j)}$ is an adiabatic correction of order $j$.  In this way we obtain an adiabatic expansion of the mode functions $h_k$ in powers of even order  adiabatic terms  ($0, 2, 4,..$.), such as $a$, $a''\propto R$, $\omega'^2$, $\omega''$, $\omega''^2$, $R^2$  etc. The non-appearance of odd adiabatic orders is justified by arguments of general covariance, which forbid tensors of odd adiabatic order  in the field equations.

\subsection{Relating different renormalization scales through the ARP}\label{sect:RelatingScales}

We start by defining the first term $\omega_k^{(0)}$ of the above WKB expansion and compute its first two derivatives:
\begin{equation}\label{omegak0}
\omega_k^{(0)} \equiv\omega_k= \sqrt{k^2+a^2 M^2}, \qquad \omega_k^\prime=a^2\mathcal{H}\frac{M^2}{\omega_k}, \qquad\omega_k^{\prime \prime}=2a^2\mathcal{H}^2\frac{M^2}{\omega_k}+a^2\mathcal{H}^\prime \frac{M^2}{\omega_k}-a^4\mathcal{H}^2\frac{M^4}{\omega_k^3}\,.
\end{equation}
Notice that in this approach the WKB expansion is performed off-shell, i.e. we use the arbitrary mass scale $M$ instead of the original mass $m$ (both being parameters of adiabatic order zero).  In this fashion the ARP can be formulated in such a way that we can relate the adiabatically renormalized theory at two scales\,\cite{Ferreiro2019}. The mass scale $M$  can play  a role similar to the scale $\mu$ in DR, but it can be given a more physical meaning.  When  $M$ is fixed at the physical mass of the quantized field ($M = m$)  we expect to obtain the renormalized theory on-shell.   By keeping the $M$-dependence we can subtract  the EMT at such value,  thus obtaining the renormalized  theory at $M$.
In the subtraction procedure, the divergences will be cancelled and the quadratic mass differences  $\Delta^2\equiv m^2-M^2$ will appear in the correction terms relating the theory at the two renormalization scales.  These differences must be reckoned as being of adiabatic order 2 since they appear in the WKB expansion together with other terms of the same adiabatic order\,\cite{Ferreiro2019}.  For  $\Delta = 0$ we recover  $M = m$ and corresponds to the usual ARP (where one renormalizes the theory only at the scale of the particle mass)\,\cite{BirrellDavies82,ParkerToms09}.
 We will use this procedure to explore the behavior of the VED throughout the cosmological evolution. The masses $m$ could be associated to fields of the Standard Model of particle physics, but as we shall see it will be convenient to consider also the heavy fields of some Grand Unified Theory (GUT) and explore the behavior in the  low energy domain $M^2\ll m^2$. Needless to say, for the sake of simplicity, we model here all particles in terms of (real) scalar fields.

We can see from Eq.\eqref{omegak0} that the expansion in terms of an even number of derivatives of $\omega_k$ (hence of even adiabatic order) is equivalent to an  expansion in even powers of $\cH$ and odd powers of $\cpH$ (notice e.g. that $\cH^2$  and $\cpH$  are homogeneous), as in both cases it involves an even number of derivatives of the scale factor. In this way the expansion is compatible with general covariance, as indicated above.  For the current universe, the powers  $\cH^2$ and $\cpH$ are sufficient for the phenomenological description, as it is obvious from the fact that  $R= (6/a^2)(\cH^2+\cpH)$,  whereas the higher powers bring corrections which can be important in the early universe.

\subsection{Computing the adiabatic orders and the regularized ZPE}\label{eq:RegZPE}

To obtain the different orders, we  start with the initial solution $W_k\approx \omega_k^{(0)}$ indicated in Eq.\eqref{omegak0}.  For  $a=1$ this would yield the standard Minkowski space modes. But since   $a=a(\eta)$ we have to find a better approximation.  Introducing that initial solution on the RHS of \eqref{WKBIteration} and expanding it in powers of  $\omega_k^{-1}$  we may collect the new terms up to adiabatic order 2 to find $\omega_k^{(2)}$. Next we iterate the procedure by introducing $W_k\approx \omega_k^{(0)}+\omega_k^{(2)}$ on the RHS of the same equation, expand again  in  $\omega_k^{-1}$   and collect the terms of adiabatic order 4, etc.  Since this mathematical procedure implies an expansion in powers of  $\omega_k^{-1}\sim 1/k\sim\lambda$ (i.e. a short wavelength expansion) it is obvious that the UV divergent terms  of the ARP are the ones containing the first lowest powers of $1/\omega_k$, and hence are concentrated in the first adiabatic orders, whilst the higher adiabatic orders represent finite contributions \cite{Bunch1980,ParkerFulling1974,BunchParker1979}.  The result is intuitive:  for any given physical quantity, the UV divergences are concentrated in the first adiabatic orders whereas  the higher orders must decay sufficiently quick at high momentum so as to make the corresponding integrals convergent and yielding a suppressed contribution.  Although not involved in our calculation, if we take  the electric current, for instance, the divergences are concentrated up to 3rd order, since here one has to include a vector potential with adiabatic order $1$.   In contrast,  for the main quantity at stake in our case,  the EMT, its regularization  implies to  work up to $4th$ adiabatic order,  as we shall show in detail below - cf. Eq.\,\eqref{EMTFluctuations}.  Upon renormalization, we will obtain a finite expression for the EMT.

Using \eqref{omegak0} and working out  the second and fourth order adiabatic terms of \eqref{WKB}, one finds
\begin{equation}
\begin{split}
\omega_k^{(2)}&= \frac{a^2 \Delta^2}{2\omega_k}+\frac{a^2 R}{2\omega_k}(\xi-1/6)-\frac{\omega_k^{\prime \prime}}{4\omega_k^2}+\frac{3\omega_k^{\prime 2}}{8\omega_k^3}\,,\\
\omega_k^{(4)}&=-\frac{1}{2\omega_k}\left(\omega_k^{(2)}\right)^2+\frac{\omega_k^{(2)}\omega_k^{\prime \prime}}{4\omega_k^3}-\frac{\omega_k^{(2)\prime\prime}}{4\omega_k^2}-\frac{3\omega_k^{(2)}\omega_k^{\prime 2}}{4\omega_k^4}+\frac{3\omega_k^\prime \omega_k^{(2)\prime}}{4\omega_k^3}\,.
\end{split}\label{WKBexpansions}
\end{equation}
We are now ready to compute the energy density associated to the quantum vacuum fluctuations in curved spacetime with FLRW metric. We start  from the EMT given in Eq.\,\eqref{EMTScalarField}  with $\phi$ decomposed as  in (\ref{ExpansionField}).  However, we are interested just on the fluctuating part, and select the quadratic fluctuations in  $\delta\phi$ only since, as previously indicated, we have zero VEV for the fluctuation itself. This follows from (\ref{FourierModes}) and the definition of adiabatic vacuum, implying that the crossed terms  $\propto\langle\delta\phi\rangle$ vanish.  For the $00$-component, related to the energy density of the vacuum fluctuations, we find
\begin{eqnarray}\label{EMTInTermsOfDeltaPhi}
\langle T_{00}^{\delta \phi}\rangle &=&\left\langle \frac{1}{2}\left(\delta\phi^{\prime}\right)^2+\left(\frac{1}{2}-2\xi\right)\sum_i\partial_i \delta \phi\partial_i \delta \phi+6\xi\mathcal{H}\delta \phi \delta \phi^\prime\right.\nonumber\\
&&\phantom{XXXXX}-\left.2\xi\delta\phi\sum_i \partial_{ii}\delta\phi+3\xi\mathcal{H}^2\delta\phi^2+\frac{a^2m^2}{2}(\delta\phi)^2 \right\rangle\,.
\end{eqnarray}
To clarify the notation, notice that $\left(\delta\phi^{\prime}\right)^2\equiv\left(\delta\partial_0\phi\right)^2= \left(\partial_0\delta\phi\right)^2$.  We may now substitute the Fourier expansion of $\delta\phi=\delta\varphi/a$, as given in \eqref{FourierModes},  into Eq.\,\eqref{EMTInTermsOfDeltaPhi} and apply the commutation relations \eqref{CommutationRelation}.
After symmetrizing the operator field product $\delta\phi \delta\phi^\prime$ with respect to the creation and annihilation operators, we end up with the following expression in terms of the amplitudes of the Fourier modes of the scalar field:
\begin{equation}
\begin{split}
&\langle T_{00}^{\delta \phi}\rangle=\frac{1}{4\pi^2 a^2}\int dk k^2 \left[ \left|h_k^\prime\right|^2+(\omega_k^2+a^2\Delta^2)\left|h_k\right|^2 \right.\\
&\left. +\left(\xi-\frac{1}{6}\right)\left(-6\mathcal{H}^2\left|h_k\right|^2+6\mathcal{H}\left(h_k^\prime h_k^{*}+h_k^{*\prime}h_k\right)\right)\right]\,,
\end{split}\label{T00}
\end{equation}
where  we have integrated $ \int\frac{d^3k}{(2\pi)^3}(...)$ over solid angles and expressed the final integration in terms of $k=|\bk|$.
The different terms of the above integral should be expanded up to $4th$ order in adiabatic expansion using the WKB approximations (\ref{WKBexpansions}):
\begin{equation}\label{eq:exphk2}
\begin{split}
&|h_k|^2=\frac{1}{2W_k}=\frac{1}{2\omega_k}-\frac{\omega_k^{(2)}}{2\omega_k^2}-\frac{\omega_k^{(4)}}{2\omega_k^2}+\frac{1}{2\omega_k}\left(\frac{\omega_k^{(2)}}{\omega_k}\right)^2+\dots\\
\end{split}
\end{equation}
\begin{equation}
\begin{split}\label{eq:exphkp2}
&|h_k^\prime|^2 =\frac{\left(W_k^\prime\right)^2}{8W_k^3}+\frac{W_k}{2}= \frac{\omega_k}{2}+\frac{\omega_k^{(2)}}{2}+\frac{\omega_k^{(4)}}{2}+\frac{1}{8\omega_k}\left(\frac{\omega_k^\prime}{\omega_k}\right)^2\left(1-3\frac{\omega_k^{(2)}}{\omega_k}\right)
+\frac{\omega_k^\prime \omega_k^{(2)\prime}}{4\omega_k^3}+\dots
\end{split}
\end{equation}
\begin{equation}
\begin{split}
h_k^\prime h_k^{*}+h_k^{*\prime}h_k &= -\frac{W_k^\prime}{2W_k^2}=-\frac{\omega_k^{\prime}}{2\omega_k^2}\left( 1-\frac{2\omega_k^{(2)}}{\omega_k}\right)-\frac{\omega_k^{(2)\prime}}{2\omega_k^2}+\dots\\
\end{split}
\end{equation}
 Upon substituting the above WKB expansions  in (\ref{T00}) and using the relations  \eqref{WKBexpansions} and \eqref{omegak0},  the result can be phrased as follows after a significant amount of algebra:
\begin{equation}
\begin{split}
\langle T_{00}^{\delta \phi} \rangle & =\frac{1}{8\pi^2 a^2}\int dk k^2 \left[ 2\omega_k+\frac{a^4M^4 \mathcal{H}^2}{4\omega_k^5}-\frac{a^4 M^4}{16 \omega_k^7}(2\mathcal{H}^{\prime\prime}\mathcal{H}-\mathcal{H}^{\prime 2}+8 \mathcal{H}^\prime \mathcal{H}^2+4\mathcal{H}^4)\right.\\
&+\frac{7a^6 M^6}{8 \omega_k^9}(\mathcal{H}^\prime \mathcal{H}^2+2\mathcal{H}^4) -\frac{105 a^8 M^8 \mathcal{H}^4}{64 \omega_k^{11}}\\
&+\left(\xi-\frac{1}{6}\right)\left(-\frac{6\mathcal{H}^2}{\omega_k}-\frac{6 a^2 M^2\mathcal{H}^2}{\omega_k^3}+\frac{a^2 M^2}{2\omega_k^5}(6\mathcal{H}^{\prime \prime}\mathcal{H}-3\mathcal{H}^{\prime 2}+12\mathcal{H}^\prime \mathcal{H}^2)\right. \\
& \left. -\frac{a^4 M^4}{8\omega_k^7}(120 \mathcal{H}^\prime \mathcal{H}^2 +210 \mathcal{H}^4)+\frac{105a^6 M^6 \mathcal{H}^4}{4\omega_k^9}\right)\\
&+\left. \left(\xi-\frac{1}{6}\right)^2\left(-\frac{1}{4\omega_k^3}(72\mathcal{H}^{\prime\prime}\mathcal{H}-36\mathcal{H}^{\prime 2}-108\mathcal{H}^4)+\frac{54a^2M^2}{\omega_k^5}(\mathcal{H}^\prime \mathcal{H}^2+\mathcal{H}^4) \right)
\right]\\
&+\frac{1}{8\pi^2 a^2} \int dk k^2 \left[  \frac{a^2\Delta^2}{\omega_k} -\frac{a^4 \Delta^4}{4\omega_k^3}+\frac{a^4 \mathcal{H}^2 M^2 \Delta^2}{2\omega_k^5}-\frac{5}{8}\frac{a^6\mathcal{H}^2 M^4\Delta^2}{\omega_k^7} \right.\\
& \left. +\left( \xi-\frac{1}{6} \right) \left(-\frac{3a^2\Delta^2 \mathcal{H}^2}{\omega_k^3}+\frac{9a^4 M^2 \Delta^2 \mathcal{H}^2}{\omega_k^5}\right)\right]+\dots, \label{EMTFluctuations}
\end{split}
\end{equation}
Let us note the presence of the $\Delta$-dependent terms in the last two rows, which contribute at second ($\Delta^2$) and fourth ($\Delta^4$) adiabatic order.

\subsection{Particular cases: ZPE with minimal coupling  and  in Minkowski spacetime} \label{sec:ParticularCases}

As a particular case of the cumbersome expression obtained above, let us consider what is left when the non-minimal coupling to gravity is absent ($\xi=0$).  Let us also fix the scale $M$ at the physical mass of the particle ($M=m$), so that the $\Delta$-terms vanish. Finally, let us project the UV-divergent terms of order $\cH^2$ and  neglect those of higher adiabatic order.  It is then easy to check that  Eq.\, \eqref{EMTFluctuations}  boils down to the very simple expression
\begin{equation}\label{eq:LowOrderZPE1}
  \left.\langle T_{00}^{\delta \phi}\rangle\right|_{M=m}=\frac{1}{8\pi^2 a^2}\int dk k^2\left( 2\omega_k(m)+\frac{\cH^2}{\omega_k(m)}+\frac{a^2m^2 \cH^2}{\omega_k^3(m)}\right)\,,
\end{equation}
where we recall that  $\omega_k(m)\equiv \sqrt{k^2+a^2 m^2}$.
Formula \eqref{eq:LowOrderZPE1} is in agreement with previous results found in the literature for $\xi=0$, in the  ${\cal O}(H^2)$ approximation\,\cite{ParkerFulling1974} --  see also\,\cite{Maggiore2011,Bilic2011}.  Notice that $k$ is the comoving momentum, whereas the physical momentum is $\tilde{k}=k/a$. Defining the physical energy mode  $\tilde\omega_k(m)=\sqrt{\tilde{k}^2+m^2}$, and keeping in mind that $\cH=a H$, we can re-express the above result as
\begin{equation}\label{eq:LowOrderZPE2}
  \left.\langle T_{00}^{\delta \phi}\rangle\right|_{M=m}=a^2\frac{1}{4\pi^2}\int d\tilde{k}\tilde{k}^2\left[\tilde\omega_k(m)+\frac{H^2}{2\tilde\omega_k(m)}\left(1+\frac{m^2}{\tilde\omega_k^2(m)}\right)\right]\,.
\end{equation}
 The dependence on the scale factor can be eliminated as soon as we write  $T_{00}=-\rho_{\rm vac}g_{00}=a^2\rho_{\rm vac}$  and rephrase the above result in terms of $\rho_{\rm vac}$. The last quantity can be thought of  as representing the VED associated to the quantum fluctuations, i.e. the aforementioned ZPE.  We will, however, come back to this point later on. At the moment we note that with this interpretation
we can  retrieve also the very particular situation of  Minkowskian spacetime, as indicated above. Setting $a=1$ (hence $H=0$)  the previous expression maximally simplifies to
\begin{equation}\label{eq:Minkoski}
  \left.\langle T_{00}^{\delta \phi}\rangle\right|_{\rm Minkowski}=\frac{1}{4\pi^2}\int dk k^2 \omega_k =  \int\frac{d^3k}{(2\pi)^3}\,\frac12\,\hbar\,\omega_k\,,
\end{equation}
where the last integral  is  just the well-known ZPE of the quantum field $\phi$ in flat spacetime\,\cite{JSPRev2013,Akhmedov2002}, as it should be expected  (in natural units,  $\hbar=1$). In what follows, unless stated otherwise,  we will continue using comoving momenta, as it will ease the presentation.
The previous formulas correspond to simpler situations, but as we have seen the ZPE in FLRW spacetime is much more complicated, and Eq.\,(\ref{EMTFluctuations}) constitutes a WKB approximation to it up to $4th$ adiabatic order.  Of course, all the above forms of ZPE are UV divergent and require renormalization.

\section{Renormalization of the ZPE in the FLRW background}\label{sec:RenormZPE}

Let us consider the ZPE part of the EMT, as given by  Eq.\,\eqref{EMTFluctuations}. We can split it into two parts as follows:
\begin{equation}
\langle T_{00}^{\delta \phi}\rangle (M)= \langle T_{00}^{\delta \phi}\rangle_{Div}(M)+\langle T_{00}^{\delta \phi}\rangle_{Non-Div}(M), \label{DecompositionEMT}
\end{equation}
where
\begin{equation}
\begin{split}
&\langle T_{00}^{\delta \phi}\rangle_{Div}(M)=\frac{1}{8\pi^2 a^2}\int dk k^2 \Bigg[ 2\omega_k +\frac{a^2 \Delta^2}{\omega_k}-\frac{a^4\Delta^4}{4\omega_k^3} -\left(\xi-\frac{1}{6}\right)6\mathcal{H}^2\left(\frac{1}{\omega_k}+\frac{a^2 M^2}{\omega_k^3}+\frac{a^2 \Delta^2}{2\omega_k^3}\right) \\
& -\left(\xi-\frac{1}{6}\right)^2\frac{9}{\omega_k^3}(2\mathcal{H}^{\prime\prime}\mathcal{H}-\mathcal{H}^{\prime 2}-3\mathcal{H}^4) \Bigg]
\end{split} \label{DivergentPart}
\end{equation}
is the  UV-divergent contribution, which involves $\omega_k =\sqrt{k^2+a^2M^2}$ and the powers  $1/\omega_k^n$ up to $n=3$.


The terms in \eqref{EMTFluctuations} which are not in \eqref{DivergentPart} are the ones which are finite  (as they  involve  powers of  $1/\omega_k$ higher than $3$),  and constitute the $\langle T_{00}^{\delta \phi}\rangle_{Non-Div}(M)$ part of  (\ref{DecompositionEMT}).  Computing the (manifestly convergent) integrals with the help of Eq.\,(\ref{DRFormula}) (for $\epsilon=0$) in  Appendix \ref{AppendixB}, the final result reads
\begin{equation}\label{Non-DivergentPart}
\begin{split}
\langle T_{00}^{\delta \phi}\rangle_{Non-Div}(M) &=\frac{m^2 \mathcal{H}^2}{96\pi^2}-\frac{1}{960\pi^2 a^2}\left( 2\mathcal{H}^{\prime \prime}\mathcal{H}-\mathcal{H}^{\prime 2}-2\mathcal{H}^4\right)\\
&+\frac{1}{16\pi^2 a^2}\left(\xi-\frac{1}{6}\right) \left(2\mathcal{H}^{\prime \prime}\mathcal{H}-\mathcal{H}^{\prime 2}-3\mathcal{H}^4\right)+\frac{9}{4\pi^2 a^2}\left(\xi-\frac{1}{6}\right)^2 (\mathcal{H}^\prime \mathcal{H}^2 +\mathcal{H}^4)\\
&+ \left(\xi-\frac{1}{6}\right)\frac{3\Delta^2 \mathcal{H}^2}{8\pi^2}+\dots
\end{split}
\end{equation}
where the dots in the last expression correspond to higher adiabatic orders.
Let us now take a closer look to the divergent part of the ZPE, Eq.\, \eqref{DivergentPart}. Since the complete adiabatic series is an asymptotic series representation of  Eq.\,\eqref{EMTInTermsOfDeltaPhi}, there is some arbitrariness in the way of choosing the leading adiabatic order because, independently of our choice, such  series does not really converge and only serves as an approximation, which is obtained after one cuts the series at  some particular order.  There is, however, a minimum number of steps to do in order to obtain a meaningful result. To start with, let us set the arbitrary scale $M$ to the on-shell mass value of the quantized scalar field,  $M=m$, hence $\Delta=0$ (cf. Sec.\ref{sect:RelatingScales}). In such a case, the divergent part (\ref{DivergentPart}) reduces to
\begin{equation}
\begin{split}
&\langle T_{00}^{\delta \phi}\rangle_{Div}(m)=\frac{1}{8\pi^2 a^2}\int dk k^2 \Bigg[ 2\omega_k(m) -\left(\xi-\frac{1}{6}\right)6\mathcal{H}^2\left(\frac{1}{\omega_k(m)}+\frac{a^2m^2}{{\omega_k^3(m)}}\right) \\
& -\left(\xi-\frac{1}{6}\right)^2\frac{9}{{\omega_k^3(m)}}(2\mathcal{H}^{\prime\prime}\mathcal{H}-\mathcal{H}^{\prime 2}-3\mathcal{H}^4) \Bigg]\,.
\end{split} \label{DivergentPartClassic}
\end{equation}
 Again, \eqref{DivergentPartClassic} is a bare integral, formally divergent and does not depend on any renormalization scale.   The prescription we are going to follow in order to renormalize the ZPE (and, in general, the EMT) is somehow reminiscent of the momentum subtraction scheme, although is  certainly different in many respects.  In the latter the renormalized Green's functions and running couplings are obtained by subtracting their values at a renormalization point $p^2=M^2$ (space-like in our metric, which becomes an Euclidean point after Wick rotation) or at the time-like one $p^2=-M^2$  (depending  on the kinematical region involved)\,\cite{Donoghue92,Manohar97}. Since  for vacuum diagrams we do not have external momenta,  here,  instead, we renormalize the ZPE by subtracting the terms that appear up to $4th$ adiabatic order at the arbitrary mass scale $M$.  This suffices  to eliminate the divergent terms through the ARP, as it is amply discussed in the literature\,\cite{BirrellDavies82,ParkerToms09,Fulling89}.

\subsection{Renormalized  ZPE off-shell}\label{sec:ZPEoffshell}

In view of the previous considerations, we will define the renormalized  ZPE in curved spacetime at the scale $M$ as follows:
\begin{eqnarray}\label{EMTRenormalized}
\langle T_{00}^{\delta \phi}\rangle_{\rm Ren}(M)&=&\langle T_{00}^{\delta \phi}\rangle(m)-\langle T_{00}^{\delta \phi}\rangle^{(0-4)}(M)\nonumber\\
&=&\langle T_{00}^{\delta \phi}\rangle_{Div}(m)-\langle T_{00}^{\delta \phi}\rangle_{Div}(M)-\left(\xi-\frac{1}{6}\right)\frac{3\Delta^2 \cH^2}{8\pi^2}+\dots,
\end{eqnarray}
where we have used the fact that $\langle T_{00}^{\delta \phi}\rangle_{Non-Div}(m)-\langle T_{00}^{\delta \phi}\rangle_{Non-Div}^{(0-4)}(M)$ yields precisely the last term of (\ref{EMTRenormalized}), as it follows immediately from Eq.\, (\ref{Non-DivergentPart}).  In these expressions,  $(0-4)$ indicates the expansion up to fourth adiabatic order and the dots in (\ref{EMTRenormalized}) denote finite terms of higher adiabatic order.  Using now Eq.\,\eqref{DivergentPartClassic}, we arrive at the result
\begin{equation}
\begin{split}
&\langle T_{00}^{\delta \phi}\rangle_{\rm Ren}(M)=\frac{1}{8\pi^2 a^2}\int dk k^2 \left[ 2 \omega_k (m)-\frac{a^2 \Delta^2}{\omega_k (M)}+\frac{a^4 \Delta^4}{{4\omega^3_k (M)}}-2 \omega_k (M)\right]\\
&-\left(\xi-\frac{1}{6}\right)6\mathcal{H}^2 \frac{1}{8\pi^2 a^2}\int dk k^2 \left[-\frac{1}{\omega_k (M)}-\frac{a^2 M^2}{{\omega^3_k (M)}}-\frac{a^2 \Delta^2}{2{\omega^3_k (M)}}+\frac{1}{\omega_k(m)}+\frac{a^2 m^2}{{\omega^3_k (m)}} \right]\\
&-\left(\xi-\frac{1}{6}\right)^2 \frac{9\left(2 \mathcal{H}^{\prime \prime}\mathcal{H}-\mathcal{H}^{\prime 2}-3 \mathcal{H}^{4}\right)}{8\pi^2 a^2}\int dk k^2 \left[ \frac{1}{{\omega^3_k (m)}}-\frac{1}{{\omega^3_k (M)}}\right]\\
&-\left(\xi-\frac{1}{6}\right)\frac{3\Delta^2  \mathcal{H}^2}{8\pi^2}+\dots
\end{split} \label{Renormalized}
\end{equation}
For better clarity, we  will henceforth distinguish explicitly  between the off-shell energy mode $\omega_k(M)=\sqrt{k^2+a^2 M^2}$  (formerly denoted just as $\omega_k$) and the on-shell one  $\omega_k(m)=\sqrt{k^2+a^2 m^2}$.
 On using simple manipulations, such as e.g.
\begin{eqnarray}\label{eq:usefuldiff}
&&\omega_k (m)-\omega_k(M)=\left(\omega_k (m)-\omega_k(M)\right)\frac{\omega_k (m)+\omega_k(M)}{\omega_k (m)+\omega_k(M)}=\frac{a^2\Delta^2}{\omega_k (m)+\omega_k(M)}\,,\nonumber\\
&& 2 (\omega_k (m)- \omega_k (M))-\frac{a^2 \Delta^2}{\omega_k (M)}+\frac{a^4 \Delta^4}{{4\omega^3_k (M)}}= \Delta^6 a^6 \frac{\omega_k (m)+3\omega_k (M)}{4\omega^3_k (M)(\omega_k (m)+\omega_k (M))^3}\,,
\end{eqnarray}
etc. one can work out the renormalized result \eqref{Renormalized} into the following convenient form
\begin{equation}
\begin{split}
&\langle T_{00}^{\delta \phi}\rangle_{\rm Ren}(M)=\frac{\Delta^6}{8\pi^2}\int_0^\infty dk k^2 a^4 \left[\frac{\omega_k (m)+3\omega_k (M)}{4\omega^3_k (M)(\omega_k (m)+\omega_k (M))^3}\right]\\
&-\left(\xi-\frac{1}{6}\right)\frac{3a^2 \mathcal{H}^2 }{4\pi^2 }\int_0^\infty dk  \Bigg[\frac{\Delta^2 m^2}{2\omega_k (m) \left( \omega_k (m)+\omega_k (m) \right)^2}+\frac{\Delta^2 M^2}{2 \omega_k (M) \left( \omega_k (m)+\omega_k (M) \right)^2}\\
&-\frac{\Delta^2 m^2}{2 \omega_k (M) \omega_k (m) \left( \omega_k (m)+\omega_k (M) \right)} -\frac{m^4}{\omega^3_k (m)}+\frac{M^4}{2\omega^3_k (M)}+\frac{M^2 m^2}{2\omega^3_k (M)}\Bigg] \\
&-\left(\xi-\frac{1}{6}\right)^2 \frac{9\left(2 \mathcal{H}^{\prime \prime}\mathcal{H}-\mathcal{H}^{\prime 2}-3 \mathcal{H}^{4}\right)}{8\pi^2 }\int_0^\infty dk  \frac{k^2}{\omega^3_k(m)\omega^3_k(M)}\left[\frac{-k^2\Delta^2}{\omega_k(m)+\omega_k(M)}+M^2 \omega_k (M)-m^2 \omega_k (m) \right]\\
&-\left(\xi-\frac{1}{6}\right)\frac{3\Delta^2  \mathcal{H}^2}{8\pi^2}+\dots\,,
\end{split} \label{RenormalizedExplicit}
\end{equation}
in which all the integrals  are seen to be manifestly convergent since the power counting for all of them leads to $\sim \int dk k^{-3}$ in the UV region.
The calculation of some of these convergent integrals can be  a bit cumbersome, as not all of them can be dealt directly with  Eq.\,(\ref{DRFormula}).   Owing to various cancellations, however, the final result can be cast in a rather compact form:
\begin{equation}
\begin{split}
&\langle T_{00}^{\delta \phi}\rangle_{\rm Ren}(M)=\frac{a^2}{128\pi^2 }\left(-M^4+4m^2M^2-3m^4+2m^4 \ln \frac{m^2}{M^2}\right)\\
&-\left(\xi-\frac{1}{6}\right)\frac{3 \mathcal{H}^2 }{16 \pi^2 }\left(m^2-M^2-m^2\ln \frac{m^2}{M^2} \right)+\left(\xi-\frac{1}{6}\right)^2 \frac{9\left(2  \mathcal{H}^{\prime \prime} \mathcal{H}- \mathcal{H}^{\prime 2}- 3  \mathcal{H}^{4}\right)}{16\pi^2 a^2}\ln \frac{m^2}{M^2}+\dots
\end{split} \label{RenormalizedExplicit2}
\end{equation}
We have checked this result with the help of  Mathematica\cite{Mathematica}.
The obtained expression vanishes for $M=m$, which was already obvious from \eqref{Renormalized} or (\ref{RenormalizedExplicit}),  since the integrand is proportional to various powers of $\Delta$ and to expressions that cancel in that limit.  This is also clear from the definition itself, Eq.\,(\ref{EMTRenormalized}).

However, it should be emphasized that the vanishing result in the $M=m$  limit  occurs only because we have computed  the on-shell value $\langle T_{\mu\nu}^{\delta \phi}\rangle_{\rm Ren}(m)$ also up to adiabatic order $4$ in Eq.\,(\ref{RenormalizedExplicit2}). In general one can compute $\langle T_{\mu\nu}^{\delta \phi}\rangle_{\rm Ren}(m)$ up to any desired adiabatic order,  keeping however in mind the asymptotic character of the WKB series. But in all cases the subtracted term in Eq.\,(\ref{EMTRenormalized})  at the arbitrary scale $M$ is always to be computed  to adiabatic order 4,  as this suffices to cancel all the existing divergences.  Beyond $4th$ order one always obtains finite,  subleading,  corrections.  These higher order finite effects satisfy the Appelquist-Carazzone decoupling theorem\,\cite{AppelquistCarazzone75} since they become suppressed  for  large values of the physical mass $m$  of the quantum field.  In our study, however, we do not track these  finite,  subleading,  contributions, but of course they are there and provide a nonvanishing on-shell value of the renormalized EMT as defined by \eqref{EMTRenormalized} .

Noteworthy, the final renormalized ZPE in curved spacetime (\ref{RenormalizedExplicit2}),  although it is perfectly finite, still carries at this point quartic powers of the masses.

We have explicitly checked that the above direct  subtraction procedure gives the same result as the conventional DR technique applied to the divergent integrals of \eqref{DivergentPart}, see  Appendix \ref{AppendixB} for  a summary of that alternative calculation. Of course, the DR is used here only as an auxiliary tool to regularize the UV divergences by tracking the poles up to adiabatic order four, but  we do not mean at all to  renormalize the calculation through the minimal subtraction (MS) scheme\,\cite{Donoghue92,Manohar97}.  In fact, as we have demonstrated, the above result can be fully obtained without any use of DR, if it is not desired. The truly guiding  renormalization principle here is the one based on the ARP relating different scales, with or without the auxiliary use of DR in the intermediate steps.

\section{Renormalized vacuum energy density}\label{sec:RenormalizedVED}

We remind the reader that in order to make possible the renormalization program in the context of QFT in curved spacetime, we need to count on the higher derivative (HD) terms  in the classical effective action of vacuum\,\cite{BirrellDavies82}, in addition to the usual Einstein-Hilbert (EH) term with a cosmological constant, $\CC$.  In four dimensions, the HD part is composed of  the ${\cal O}(R^2)$ terms, i.e. the squares of the curvature and Ricci tensors:  $R^2$ and $R_{\mu\nu}R^{\mu\nu}$. No more HD terms are needed in our case since the one associated to the square of the Riemann tensor, $R_{\mu\nu\rho\sigma}R^{\mu\nu\rho\sigma}$, is not independent owing to the topological nature of the Euler's density in $4$ dimensions, which involves all these  HD terms together: $E=R_{\mu\nu\rho\sigma}R^{\mu\nu\rho\sigma}-4R_{\mu\nu}R^{\mu\nu}+R^2$.  Moreover the square of the Weyl tensor, $C^2=R_{\mu\nu\rho\sigma}R^{\mu\nu\rho\sigma}-2R_{\mu\nu}R^{\mu\nu}+(1/3)\,R^2$, exactly vanishes for conformally flat spacetimes such as FLRW.  The full action, therefore, boils down to the relevant EH+HD terms mentioned above plus the matter part (consisting here  of the scalar field $\phi$ only) with a non-minimal coupling to gravity, Eq.\,(\ref{eq:Sphi}).  Variation of the action with respect to the metric provides the modified Einstein's equations, which become extended as compared to their original form \eqref{FieldEq2} as follows:
\begin{equation}\label{eq:MEEs}
\frac{1}{8\pi G_N(M)}G_{\mu \nu}+\rho_\Lambda(M) g_{\mu \nu}+a_1(M) H_{\mu \nu}^{(1)}= T_{\mu \nu}^{\phi_b} +\langle T_{\mu \nu}^{\delta \phi} \rangle_{\rm Ren}(M)\,,
\end{equation}
where we use renormalized quantities and hence we have indicated explicitly the dependence of the parameters and of the  EMT on the subtraction point  $M$.  The background part does not depend on it.
The higher order tensor  $H_{\mu \nu}^{(1)}$  is  obtained  by functionally differentiating  $R^2$
with respect to the metric (see Appendix \ref{AppendixA}).  A further simplification is possible here since the corresponding term associated to the functional differentiation of the square of the Ricci tensor, called  $H_{\mu \nu}^{(2)}$, is not necessary since it is not independent of $H_{\mu \nu}^{(1)}$  for  FLRW spacetimes\,\cite{BirrellDavies82}.  This follows from the aforementioned properties of the Euler density and the Weyl tensor for conformally flat spacetimes.  The higher order tensor $H_{\mu \nu}^{(1)}$ is indeed to be included in the extended  field equations since it is anyway generated by the quantum fluctuations and is therefore indispensable for the renormalizability of the theory.
The fact that  Eq.\,\eqref{eq:MEEs} has been written with all couplings defined at some arbitrary mass scale $M$ is because we have shown that the EMT used in our calculation is the renormalized one at that scale following the ARP.  However, in the Appendix \ref{AppendixB} we offer an alternative approach based on the more conventional counterterm procedure, starting from the bare parameters of the action.

Baring in mind that we wish to relate the  theory at different renormalization points\footnote{Renormalization theory is concerned with the relations of renormalized couplings, operators and Green's functions at different renormalization points. It is not our intention to compute any of these quantities from first principles, in particular the VED. Ultimately this is an input from experiment at a given scale, and once it is given one can predict its value at another scale. }, let us subtract the modified Einstein's equations \eqref{eq:MEEs} at the two scales $M$ and $M_0$.  The classical (background) contribution $T_{\mu \nu}^{\phi_b}$ cancels in the difference, since as noted it does not depend on the renormalization scale, and we  find
\begin{equation}
\langle T_{\mu \nu}^{\delta \phi}\rangle_{\rm Ren}(M)- \langle T_{\mu \nu}^{\delta \phi}\rangle_{\rm Ren}(M_0)=f_{G_N^{-1}}(m,M,M_0)  G_{\mu \nu}+f_{\rho_\CC}(m,M,M_0) g_{\mu \nu}+f_{a_1}(m,M,M_0) H^{(1)}_{\mu \nu},  	\label{EinsteinDifferentScale}
\end{equation}
where we have introduced the subtracted parameters
\begin{equation}\label{eq:SubtractedX}
f_X(m,M,M_0)\equiv X(M)- X(M_0)
\end{equation}
for the various couplings involved $X=\joantext{G_N^{-1}/(8\pi)}, \rL, a_1$. \joantext{(For simplicity, we denote $f_{G_N^{-1}/8\pi}$ just as $f_{G_N^{-1}}$.})
Using  now the tensor pattern shown by the generalized field equations \eqref{eq:MEEs}, and taking into account that we know the expression for the final renormalized form of the  EMT within the ARP, namely  Eq\, (\ref{RenormalizedExplicit2}),  we can derive by comparison the renormalization shift (or  `running')  undergone by the couplings  $G_N^{-1}$, $\rL$  and $a_1$ in \eqref{EinsteinDifferentScale} between the two scales $M$ and $M_0$.  Such identification is possible since  we know the explicit expressions for  $G_{00}$ and $H_{00}^{(1)}$ -- see Appendix \ref{AppendixA}.  The former is proportional to $\cH^2$ (adiabatic order $2$)  and the latter to a linear combination of  terms of adiabatic order $4$ involving $\cH$ and its derivatives -- cf. Eqs.\,\eqref{eq:R}, \eqref{eq:R00G00} and \eqref{eq:H100}.  The remaining term of (\ref{RenormalizedExplicit2}) -- the first one on its \textit{r.h.s} -- is of adiabatic order zero; it  is associated to the running  of $\rL$ and determines $f_{\rho_\CC}(m,M,M_0)$.
Explicitly, setting $\mu=\nu=0$ we find the results
\begin{equation} \label{SubtractionOfG}
f_{G_N^{-1}}(m,M,M_0)=\left(\xi-\frac{1}{6}\right)\frac{1}{16\pi^2}\left[M^2 - M_0^{2} -m^2\ln \frac{M^{2}}{M_0^2}\right],
\end{equation}
\begin{equation} \label{SubtractionOfRhoL}
f_{\rho_\CC}(m,M,M_0) =\frac{1}{128\pi^2}\left(M^4-M_0^{4}-4m^2(M^2-M_0^{2})+2m^4\ln  \frac{M^{2}}{M_0^2}\right),
\end{equation}
and
\begin{equation} \label{SubtractionOfa1}
f_{a_1}(m,M,M_0)= \frac{1}{32\pi^2}\left(\xi-\frac{1}{6}\right)^2 \ln \frac{M^2}{M_0^{2}}.
\end{equation}

\subsection{Vacuum energy density at different scales. Absence of $\sim m^4$ terms.}\label{sec:TotalVED}

Following our discussion in Sec.\,\ref{eq:RegZPE}, let us provisionally define  the   vacuum state  as that one satisfying $p_{\rm vac}=- \rho_{\rm vac}$  and  $ T_{\mu \nu}^{\rm vac} =-\rho_{\rm vac} g_{\mu\nu}$.  We will further discuss the significance of this identification in Appendix \ref{AppendixC}.  Equating the last expression to  Eq.\,\eqref{EMTvacuum},
and taking the $00$-component of the equality (keeping also in mind that $g_{00}=-a^2(\eta)$ in the conformal frame), we obtain
\begin{equation}\label{eq:Totalrhovac}
\rho_{vac}(M)=\rho_{\Lambda}(M)+\frac{\langle T_{00}^{\delta \phi}\rangle_{\rm Ren}(M )}{a^2}.
\end{equation}
Notice that we have included the dependence on the renormalization point since we are using the renormalized theory at that scale.
The above equation says that the total VED at an arbitrary  scale $M$  is the sum of the renormalized contributions from the cosmological term plus  that  of the quantum fluctuations of the scalar field at that scale  (i.e. the renormalized ZPE).  Subtracting the renormalized result at two scales, $M$ and $M_0$, and using \eqref{EinsteinDifferentScale}, we find:
\begin{equation}
\begin{split}
&\rho_{vac}(M)-\rho_{vac}(M_0)=\rho_{\CC}(M)-\rho_{\CC}(M_0)+\frac{\langle T_{00}^{\delta \phi}\rangle_{\rm Ren}(M)- \langle T_{00}^{\delta \phi}\rangle_{\rm Ren}(M_0)}{a^2}\nonumber\\
&=f_{\rho_\CC}(m,M,M_0)+\frac{f_{G_N^{-1}}(m,M,M_0)  G_{00}+f_{\rho_\CC}(m,M,M_0) g_{00}+f_{a_1}(m,M,M_0) H^{(1)}_{00}}{a^2}\nonumber\\
&=\frac{f_{G_N^{-1}}(m,M,M_0) }{a^2}  G_{00}+\frac{f_{a_1}(m,M,M_0)}{a^2} H^{(1)}_{00}\nonumber
\end{split}
\end{equation}
\begin{equation}
\phantom{xx}=\frac{3\cH^2}{a^2}\,f_{G_N^{-1}}(m,M,M_0)-\frac{18}{a^4}\left(\mathcal{H}^{\prime 2}-2\mathcal{H}^{\prime \prime}\mathcal{H}+3 \mathcal{H}^4 \right)\,f_{a_1}(m,M,M_0)\,,\phantom{\,\,\,aaaaaaaaa}\label{RenormalizedVEa}
\end{equation}
where the  term $f_{\rho_\CC}(m,M,M_0)$ has cancelled, and we have used the expressions for $G_{00}$ and $H^{(1)}_{00}$ given in the Appendix \ref{AppendixA}.  From equations  (\ref{SubtractionOfG}) and (\ref{SubtractionOfa1},) we finally obtain
\begin{equation}
\begin{split}
\rho_{vac}(M)
=&\rho_{vac}(M_0)+\frac{3}{16\pi^2}\left(\xi-\frac{1}{6}\right) H^2\left[M^2 - M_0^{2}-m^2\ln \frac{M^{2}}{M_0^2}\right]\\
&-\frac{9}{16\pi^2} \left( \xi-\frac{1}{6}\right)^2 \left(\dot{H}^2 - 2 H\ddot{H} - 6 H^2 \dot{H} \right)\ln \frac{M^2}{M_0^2}\,, \label{RenormalizedVE}
\end{split}
\end{equation}
where, in addition, we have used Eq.\,(\ref{eq:H100}) to re-express the final result in terms of the ordinary Hubble function in cosmic time ($H=\cH/a$) since it will be useful for further considerations.
The  result \eqref{RenormalizedVE} is  the value of the VED  at the scale $M$  once we know its value at another  scale $M_0$, i.e. it  expresses the `running' of the VED.  Only in the case of conformally invariant fields ($\xi=1/6$) the result would be the same at all scales, if the VED would receive only contributions from scalar fields. But in general, this is not the case since one has to add the contribution from fermions and vector boson fields, which we do not consider here, so in general the total VED appears a running quantity with the expansion. The running  is slow for small $H$, as it depends on terms  of the form ${\cal O}(H^2)$ times a mass scale squared,  and on ${\cal O}(H^4)$ contributions, but not on quartic mass scales.

\subsection{Equivalent approach: subtracting the Minkowskian contribution}\label{Sec:SubtractMinkowski}

It cannot be overstated that the above result \eqref{RenormalizedVE} is free from quartic powers of the masses. These would still be present if we had subtracted  just the ZPE part at different scales without including the renormalized $\rL$. This is obvious from Eq.\eqref{RenormalizedExplicit2}, where we can see that the problem actually stems from Minkowskian spacetime, see \cite{JSPRev2013} for a discussion.  The renormalized ZPE in flat spacetime is obtained from Eq.\eqref{RenormalizedExplicit2} in the limit $a=1$ (which implies that $\mathcal{H}$ and all its derivatives are zero). Only the first term of it remains, although it  is the one  carrying the mentioned quartic powers.  This term vanishes for $M=m$  since the renormalized on-shell value was  computed only up to fourth adiabatic order.  As previously emphasized  (cf. Sec.\,\ref{sec:ZPEoffshell}),  this does not mean that the exact  renormalized ZPE vanishes on-shell, of course.  One still has to add  the higher order adiabatic terms, but they are finite and subleading since they decouple for larger and larger values of the physical mass $m$ (i.e. they satisfy the decoupling theorem\,\cite{AppelquistCarazzone75}), and we have not tracked them explicitly.  Our main aim here was to pick out just the leading contributions to the renormalized ZPE up to $4th$ adiabatic order.

Because we compute  the total VED, defined as the sum of the renormalized value of $\rL$ and the  renormalized ZPE,  the difference of VED values  at two scales is free from the quartic powers of mass scales. Of course this is possible owing to the renormalized form for the ZPE that we have used,  Eq.\,\eqref{EMTRenormalized}, which involves a subtraction of the on-shell value at another arbitrary mass scale.  In the Appendix\,\ref{AppendixB}, we provide an alternative calculation leading to the same result \eqref{RenormalizedVE} and further comments on this fact.

The above observations suggests that we can recover the expression \eqref{RenormalizedVE} for the VED  by performing an analogy with the Casimir effect; that is to say,  we may compute the expression for $\langle T_{00}^{\delta \phi}\rangle$ in Minkowskian spacetime and  substract it from its equivalent in curved spacetime.  One should expect that the result appears only mildly evolving with the cosmic evolution through a function of the Hubble rate (which is the key term providing the departure of the FLRW background from Minkowskian spacetime)\,\cite{JSPRev2013}. In fact, the subtraction of the Minkowskian spacetime result has been argued from different perspectives\,\cite{Maggiore2011,Bilic2011}.  In  the Minkowski limit,  the subtraction of scales in Eq.\,(\ref{EinsteinDifferentScale}) leaves only the term  $f_{\rho_\CC}(m,M,M_0) g_{\mu \nu}=f_{\rho_\CC}(m,M,M_0)\eta_{\mu\nu}$ on its \textit{r.h.s.} Taking the $00$-component (with  $\eta_{00}=-1$ in our conventions), we find
\begin{equation}\label{eq:MinkowskiSubtracted}
\langle T_{00}^{\delta \phi}\rangle_{\rm Ren}^{Mink}(M)-\langle T_{00}^{\delta \phi}\rangle_{\rm Ren}^{Mink}(M_0)= -f_{\rho_\CC}(m,M,M_0)\,.
\end{equation}
Following the above proposal, we define now the physical VED in the expanding universe as the outcome of subtracting  the Minkowskian ZPE  from its value in  FLRW spacetime:
\begin{equation}
\begin{split}
&\rho_{vac}(M)\equiv\frac{\langle T_{00}^{\delta \phi}\rangle_{\rm Ren}(M )}{a^2}-\left[\frac{\langle T_{00}^{\delta \phi}\rangle_{\rm Ren}(M )}{a^2}\right]^{Mink}
=\frac{\langle T_{00}^{\delta \phi}\rangle_{\rm Ren}(M )}{a^2}-\langle T_{00}^{\delta \phi}\rangle_{\rm Ren}^{Mink}(M)\,.
\end{split}
\end{equation}
Thus, inserting  equations \eqref{EinsteinDifferentScale} and (\ref{eq:MinkowskiSubtracted}) in the above relation and recalling again that $g_{00}=-a^2$, we are led to
\begin{equation}
\begin{split}
\rho_{vac}(M)&=\frac{\langle T_{00}^{\delta \phi}\rangle_{\rm Ren}(M_0)}{a^2}-\langle T_{00}^{\delta \phi}\rangle_{\rm Ren}^{Mink}(M_0)\nonumber\\
&+\frac{f_{\rho_\CC}(m,M,M_0)}{a^2} g_{00}+\frac{f_{G_N^{-1}}(m,M,M_0)}{a^2}  G_{00}+\frac{f_{a_1}(m,M,M_0)}{a^2} H^{(1)}_{00} + f_{\rho_\CC}(m,M,M_0)\nonumber
\end{split}
\end{equation}
\begin{equation}\label{AlternativeDef}
\phantom{XXXx\,x}=\rho_{vac}(M_0)+\frac{3\cH^2}{a^2}\,f_{G_N^{-1}}(m,M,M_0)-\frac{18}{a^4}\left(\mathcal{H}^{\prime 2}-2\mathcal{H}^{\prime \prime}\mathcal{H}+3 \mathcal{H}^4 \right)\,f_{a_1}(m,M,M_0)\,. \phantom{xx}
\end{equation}
The result is indeed the  same as in Eq.\,\eqref{RenormalizedVEa}, and hence we end up once more with the formula\,\eqref{RenormalizedVE} for the total VED  after we cast $\cH$ and its derivatives in terms of the ordinary Hubble rate, $H$.  In other words, we can reach again the same relation between the values of VED at two different scales, which does not involve $\sim m^4$ contributions.

Remember that the divergences associated to our calculation are of course of UV type,  hence short-distance effects.  The leading effects of this kind are similar to the ones of QFT in  Minkowski spacetime and therefore are independent from the possible boundary effects of the cosmological spacetime.  We have just seen that an alternative way to renormalize the energy-momentum tensor is precisely to subtract the Minkowskian contribution following the adiabatic regularization procedure up to fourth order.  Furthermore,  if one takes into account only wavelengths under the horizon (i.e. for $\tilde{k}^2\gg H^2$, with $\tilde{k}=k/a$ the physical momentum defined in Sec. \ref{sec:ParticularCases}), the situation remains  as in the Minkowskian spacetime, namely  the integrals with low inverse powers of the function $\omega_k=\sqrt{k^2+a^2 M^2}$, corresponding to the lowest adiabatic orders, are still divergent in the UV.  The short-distance region where the UV effects are encountered is of course contained within the horizon.  The presence of a causal horizon can only produce long distance effects, and therefore they can  be related with IR (infrared) divergences. The IR behavior of gravity theories can indeed be nontrivial in some cases but we do not address these aspects in our work as they are out of its scope. However, if we would consider effects of this kind in our momentum  integrals they would rather be related with the lower limits of integration, which should be of order $H$,  since the (apparent) horizon is of order $1/H$ (in fact, it is exactly so in the spatially flat case, which we are considering)  and the effects that could produce are subleading. To see this,  take for instance the simple cases analyzed in Sec. \ref{sec:ParticularCases}, say Eq.\,\eqref{eq:LowOrderZPE2}.  Since the physical momenta satisfy $\tilde{k}^2\ll m^2$ in the IR, the contribution from these integrals in the IR region provides powers of $H$ higher than $H^2$  involving also masses, e.g. $m H^3$, $H^5/m$ etc.  Similar terms carrying suppressed powers would appear if the more complete expression \eqref{EMTFluctuations} would be used.  The presence of odd powers of $H$  is not surprising since we have put boundaries to an otherwise covariant integration.

\section{Running vacuum connection}\label{sec:RunningConnection}

As previously remarked in a footnote,  the result we were aiming at and which is represented now by Eq.\,\eqref{RenormalizedVE} does not provide the calculated value of the vacuum energy density at a given scale, e.g. it says nothing on the value of $\rho_{vac}(M_0)$ and hence it  has no implication on the cosmological constant problem mentioned in the Introduction. That is to say, it has no bearing on it  if such problem is meant to be the computation of the value itself of the VED at some point in the history of the universe.  However, our result can  be useful to explore the `running' of the VED when we move from one scale to another. In other words, if $\rho_{vac}$ is known at some scale $M_0$, we can use the obtained relation to compute the value of $\rho_{vac}$ at another scale $M$. Such connection of values from one renormalization point to another is what we have been calling ``running'' of the VED, and in fact it was suggested long ago from the point of view of the renormalization group in curved spacetime from different perspectives\,\cite{ShapSol1,Fossil2008,ShapSol2} -- for a review of the running vacuum model (RVM), see  \cite{JSPRev2013,JSPRev2015,AdriaPhD2017} and references therein. Interestingly enough, it can provide also a framework for the possible time variation of the so-called fundamental constants of nature\,\cite{FritzschSola}.

Let us mention that different extensions of gravity can mimic the effective behavior of the running vacuum model. This  is a fact confirmed in a variety of contexts.  For instance,   in the context of Brans-Dicke Theory with a cosmological term, it has been shown that a kind of RVM behavior emerges when one tries to rewrite the theory in a GR-like picture\,\cite{GRF2018,JavierMPLA2018}.  This turns out to be phenomenologically very favorable, as it has been recently demonstrated from detailed analyses where the model has been confronted with a large and updated set of cosmological observations\,\cite{ApJL2019,BD2020}.  Another potentially interesting example can be found in gravity theories with torsion, see e.g.\,\cite{Cai2015} and references therein.  Since the torsion scalar $T$  differs only by a total derivative with respect to the Ricci scalar, the EH action with $R$ replaced by $T$ is equivalent to  GR.  One may generalize  the action structure through the replacement  $T\rightarrow T+f(T)$,  with  $f(T)$ a function  of the torsion scalar.  This is characteristic of  teleparallel gravity theories\,\cite{Cai2015}.  Since  $T=-6H^2$ in the FLRW background, by an appropriate choice of $f(T)$ one may, in principle, mimic the RVM as well.  In Sec.\,\ref{sec:RVMinflation} we discuss another example, in this case  in the context of the low-energy effective action of string theory,  which also behaves as the RVM.

Let us now come back to the obtained expression for the VDE.  In order to illustrate a  possible interpretation of  Eq.\,\eqref{RenormalizedVE} along the lines of the RVM, let us assume that we define the renormalized VED at some Grand Unified Theory (GUT) scale  $M_0=M_X$, where typically  $M_X\sim 10^{16}$ GeV is associated also with the inflationary scale.  It is natural to assume that the fundamental parameters of cosmology, such as e.g.  $\rv$,   are  primarily defined at that scale, which appeared from the very beginning in the history of the universe. By choosing a GUT scale we also insure that most matter fields can be active degrees of freedom to some extent.

\subsection{RVM in the current universe}\label{Sec:RVMCurrentUniverse}

Let  $\rv(M_X)$ be the value of the  VED at the GUT scale $M_0=M_X$.  Despite the fact that  $\rv(M_X)$  is unknown,  it can be related to the current value of the VED, $\rvo$,  through the relation $\rv(M=H_0)=\rvo$,  in which we choose  the second scale $M$ at  today's numerical value of the Hubble parameter, $H_0$.  This quantity can be used as an estimate for the energy scale of the background gravitational field associated to the FLRW universe at present.  Notice that this is precisely the kind of association originally made in the aforementioned references on the RVM\,\cite{JSPRev2013}.  Therefore, from \eqref{RenormalizedVE} applied to the current universe, we find the connection between the vacuum densities at the two points:
\begin{equation}\label{eq:rlMX1}
\rvo=\rho_{vac}(M_X)+\frac{3}{16\pi^2}\left(\frac{1}{6}-\xi\right) H_0^2\left[M_X^2+{m^2}\ln \frac{H_0^{2}}{M_X^2}\right]\\
\end{equation}
where we have neglected all terms of order  ${\cal O}(H^4)$ (which include also  $\dot{H}^2, H\ddot{H}$ and  $H^2 \dot{H}$)  for the present universe ($H=H_0$). This equation can be used to find out the unknown value of $\rho_{vac}(M_X)$, and can be conveniently written as
\begin{equation}\label{eq:rlMX2}
\rho_{vac}(M_X)=\rvo-\frac{3\nueff}{8\pi}\,H_0^2\,M_P^2\,,
\end{equation}
where we have defined the `running  parameter' for the VED:
\begin{equation}\label{eq:nueff}
\nueff=\frac{1}{2\pi}\,\left(\frac{1}{6}-\xi\right)\,\frac{M_X^2}{M_P^2}\left(1+\frac{m^2}{M_X^2}\ln \frac{H_0^{2}}{M_X^2}\right)\,.
\end{equation}
For the particular value $\xi=0$ and $m^2/M_X^2\ll1$,  the above parameter boils down to $\nueff\simeq\frac{1}{12\pi}\,\frac{M_X^2}{M_P^2}\ll1$.  Under similar conditions, but for $\xi\neq0$, the sign of $\nueff$ depends entirely on the value of $\xi$ (if only a scalar field would contribute).  As we can see,  by keeping  $\xi\neq0$ we can provide  a discussion within a  more general class of theories and also carrying potential phenomenological consequences.  At the same time it allows to confirm the expected fact that for $\xi=1/6$ there are no corrections to the vacuum energy density from scalars since we are then in the conformal limit of QFT.   Ultimately,  the final sign of $\nueff$ has to be determined by fitting the model to data. As indicated in Sec. \ref{sec:TotalVED}, this would not automatically determine the sign of $\xi$, though, since other contributions (e.g. from fermion fields) should be added in our calculation, which we leave for an independent study.  However, we understand that the basic facts derived from the renormalization procedure followed here should also hold in the general case.

Remarkably, for general  $\xi$  the structure  obtained for  $\nueff$  is very close to that obtained within the RVM approach, see \cite{JSPRev2013}.  In such context, it defines the coefficient of the one-loop $\beta$-function for the renormalization group equation of $\rho_{vac}$.  The presence of the additional logarithmic piece $\ln {H_0^{2}}/{M_X^2}$  appears in   the direct QFT calculation employed here, but it does not make any difference in practice since it  is constant and $\nueff$ must be fitted directly to the observations as an effective coefficient.  In our case we have simplified the theoretical calculation  by considering just the contribution from one single scalar field to $\nueff$. We expect it to be small, i.e. $|\nueff|\ll1$,  owing to the ratio $M_X^2/M_P^2\sim 10^{-6}$.  However the final value could be much larger  since $\nueff$ depends on $\xi$ and also  on the multiplicity and nature (fermion/boson) of the fields involved, so we cannot predict  $\nueff$ with precision on mere theoretical grounds.  It must be confronted against observations. Notice that the standard model particles make no significant contribution, since for all of them $m^2/M_X^2\ll1$. Only particles near or of order of the GUT scale may contribute significantly.  The accurate determination of  $\nueff$  can only be obtained by fitting the RVM to the overall cosmological data, as it has been done in detail e.g. in \cite{RVMpheno1,RVMpheno2}, where it has been found to be positive and of order $10^{-3}$.

Substituting\,\eqref{eq:rlMX2} into  Eq.\,\eqref{RenormalizedVE} and limiting ourselves once more to the late universe (where all terms of ${\cal O}(H^4)$ can be neglected), we can estimate the VED near our time by  taking  $M$ of order of the energy scale defined by the numerical value of $H$ around the current epoch:
\begin{equation}\label{eq:RVM1}
\rho_{vac}(H)=\rvo-\frac{3\nueff}{8\pi}\,H_0^2\,M_P^2+\frac{3\nueff(H)}{8\pi}\,H^2\,M_P^2\,,
\end{equation}
where
\begin{equation}\label{eq:nueffH}
\nueff(H)=\frac{1}{2\pi}\,\left(\frac{1}{6}-\xi\right)\,\frac{M_X^2}{M_P^2}\left(1+\frac{m^2}{M_X^2}\ln \frac{H^{2}}{M_X^2}\right)\,.
\end{equation}
Mind that the last expression depends on $H$  whereas \eqref{eq:nueff} is constant. However, being  the time evolution of $\nueff(H)$  logarithmic, and for values of $H$ not very far away from $H_0$, we can approximate $\nueff(H)$ by \,\eqref{eq:nueff}.  Then, equation (\ref{eq:RVM1}) may be cast in the more compact form
\begin{equation}\label{eq:RVM2}
\rho_{vac}(H)\simeq \rvo+\frac{3\nueff}{8\pi}\,(H^2-H_0^2)\,M_P^2=\rvo+\frac{3\nueff}{8\pi G_N}\,(H^2-H_0^2)\,,
\end{equation}
which matches the exact canonical form of the  RVM formula\,\cite{JSPRev2013}.  Let us note that such approximation holds reasonably well  even if we explore the CMB epoch since the departure of $\nueff(H)$ from $\nueff$ is  less than $8\%$ (for $m\simeq M_X$) or much less if $m\ll M_X$.

As we can see from Eq.\eqref{eq:RVM2},  for  $\nueff>0$ the vacuum can be conceived as decaying into matter since the vacuum energy density is larger in the past (where $H>H_0$), whereas if $\nueff<0$ the opposit occurs.  The former  situation, however, is more natural from a thermodynamical point of view, for if the vacuum decays into matter  one can show that the Second Law of  Thermodynamics is satisfied by the general RVM, see \cite{Yu2020} for a detailed discussion.   Moreover, for $\nueff>0$  the RVM effectively behaves as quintessence since the vacuum energy density decreases with time. One may also interpret here that $G_N$ is changing with time owing to vacuum decay. Both possibilities have been discussed  within the RVM in Ref.\,\cite{FritzschSola}. Recall that we expect  $|\nueff|\ll1$  from the theoretical structure \eqref{eq:nueffH} and, remarkably enough, we confirm it from the phenomenological fits\,\cite{RVMpheno1,RVMpheno2}, whereby we do not observe dramatic deviations from the standard $\CC$CDM model.  But the fact that the fitting results point to $\nueff= +{\cal O}(10^{-3})$ suggests that the effects are not necessarily negligible and in fact they can be helpful to cure or alleviate some of the existing tensions in the context of the $\CC$CDM model, as actually shown in the aforementioned references and also in the framework of alternative cosmological models which also mimic the RVM behavior\,\cite{ApJL2019,Mehdi2019,BD2020}.

The above equation for the VDE is the one which has been used to fit the value of $\nueff$ (assumed constant) in a variety of works, such as e.g. \cite{RVMpheno1,RVMpheno2}. As a matter of fact, such works have considered a more general form as well, in which a term proportional to $\dot{H}$ is also present in the running equation for the VED.  Such term can appear under conditions that are discussed in Appendix \ref{AppendixC}.

\subsection{Implications for the early universe: RVM-inflation}\label{sec:RVMinflation}

So far we have elaborated on the VED expression \eqref{RenormalizedVE} in the low energy regime, in which we can neglect the ${\cal O}(H^4)$ terms of the form $\dot{H}^2, H\ddot{H}$ and  $H^2 \dot{H}$.  In such regime we know that the VDE can be put in the alternative form \eqref{eq:RVM2}, which fits in with the traditional RVM structure of the vacuum evolution and represents a small dynamical departure  with respect to the $\Lambda$CDM since $|\nueff|\ll1$. Rephrased in this fashion we can see that the obtained VED around our time represents a small variation with respect to the current value of the vacuum energy density, $\rho_{\rm vac}^0$. While the previous discussion obviously applies to the current universe only, since we have neglected the ${\cal O}(H^4)$ terms on the \textit{r.h.s.} of Eq.\,\eqref{RenormalizedVE}, we should emphasize that these terms can play a significant role in the early universe. They are  generated from the functional differentiation of the $R^2$-term in the higher derivative part of the vacuum action (cf.  Appendix \ref{AppendixA}), and therefore they play a similar role as in the case of Starobinsky's inflation\,\cite{Starobinsky80,Vilenkin85}.  Notice that even though  all the terms  of the form $\dot{H}^2, H\ddot{H}$ and  $H^2 \dot{H}$ in Eq.\,\eqref{RenormalizedVE}  are denoted here as being of  ${\cal O}(H^4)$,  none of them is really proportional  to $ H^4$.  As a result, they all vanish for $H$ strictly constant.  In fact, Starobinsky's inflation is not triggered by an early epoch in which $H=$const.  but by one in which $H$ decreases at constant rate $\dot{H}=$const, see e.g. \cite{JSPRev2015} for a summarized discussion focusing on these well-known facts.  The corresponding inflation period  is characterized by a final phase with rapid oscillations of the gravitational field, which is when the universe leaves the inflationary phase and enters the radiation epoch after a reheating period. Prior to the oscillatory phase, hence within the inflationary period, $H$ decreases fast and $\dot{H}$ remains approximately constant (thence $\ddot{H}\simeq 0$)\cite{JSPRev2015}. It follows that the dominant terms  in Eq.\,\eqref{RenormalizedVE} among the  ${\cal O}(H^4)$ ones  are  $\dot{H}^2$ and $- 6 H^2 \dot{H} $, both being positive ($\dot{H}<0$).  Furthermore, since $M_0\simeq M_X$ is a higher scale (typically a GUT scale) where the couplings are defined and $M$ is some scale below it, the log term is negative. Finally, taking into account the overall minus sign in that expression (irrespective of the value of $\xi$)  we conclude that the leading contribution from the ${\cal O}(H^4)$ terms in the relevant period is positive.  After the inflation period is accomplished we know that the universe enters the radiation epoch, where $R=0$.  Henceforth these terms become irrelevant for the driving of the cosmic expansion.  We conclude that in all of the  relevant situations of the cosmic history, whether in the early universe or the late universe,  the formula \eqref{RenormalizedVE} provides a well-defined and positive expression for the evolution of the vacuum energy density.

All that said,  there are features of the RVM in the very early universe which our analysis (strictly based on QFT) could not be sensitive to, and hence we would like to comment on them here.  These features are connected with string theory contributions. In contrast to the Starobinsky form of the higher order terms mentioned above (all of which vanish for $H=$const.), the  effective generation of terms proportional to $H^4$ in the early universe  is perfectly possible from string-inspired mechanisms, see \cite{BMS2020A,BMS2020B}, in which the $\sim H^4$ power is generated in the early universe  from the vacuum average of the (anomalous) gravitational Chern-Simons term  $\sim M_P\alpha' b(x) R_{\mu\nu\rho\sigma}(x)\, \widetilde R^{\mu\nu\rho\sigma}(x)$,  which is characteristic  of the  bosonic part of the low-energy effective action of the string gravitational multiplet.  Here $b(x)$ is the Kalb-Ramond axion field and  $\alpha'$ is the slope parameter ($M_s=\sqrt{1/\alpha'}$ being the  string scale).  An effective metastable  vacuum is conceivable in this context since such state can be sustained until the universe transits into the radiation phase, and this  occurs only after the gravitational anomalies are cancelled.  This must indeed happen because matter (relativistic and nonrelativistic particles) cannot coexist with gravitational anomalies.  These can actually be cancelled  by the chiral anomalies of matter itself, see\,\cite{BMS2020A,BMS2020B} for details.  Before such thing occurs, a metastable de Sitter period remains temporally active and can bring about inflation through the (anomaly-generated)  $H^4$ term. The type of inflation produced by the $H^4$-term --- and, in general, by  higher order  (even) powers of $H$ --- is characteristic of  RVM-inflation. The latter  follows a different pattern as compared to Starobinsky's inflation, but graceful exit is still granted -- see\,\cite{rvmInflation,GRF2015,Yu2020} for details and particularly \cite{JSPRev2015} for a comparison with Starobinsky's inflation\footnote{A detailed study of $H^4$-induced (and, in general, $H^{2n}$-induced) inflation and related considerations concerning cosmological horizons and entropy can be found in \cite{Yu2020}.}.

It seems clear that the presence of the higher powers of the Hubble rate  in the early universe can be very important from different perspectives. For example, as noted in \,\cite{BMS2020A},  they could help eschewing the possible trouble of string theories with the  `swampland' criteria on the  impossibility to construct metastable de Sitter vacua in the string framework\,\cite{Vafa}, which if so it would forbid the existence of de Sitter solutions in a low energy effective theory of  quantum gravity.  The existence of the $H^4$- terms does not depend on picking out a particular potential for the scalar field since, as we should recall here,  no potential  has been introduced at any time in our framework nor in that of\,\cite{BMS2020A,BMS2020B}. Thus, the RVM string inflation approach could  provide a loophole to the swampland no-go criterion applied to fundamental scalar fields. But, of course, to fully establish it requires of a detailed investigation in the context of string-induced RVM\,\cite{BMS2020A,BMS2020B},
which is certainly not the subject of the present paper.  What is phenomenologically relevant, though, is that once these terms are available they can be used to build up a generalized form of VED, which reads as follows:
\begin{equation}\label{eq:rhoLambdaunified}
\rho_{\Lambda}(H) = \frac{3}{8\pi G_N}\left(c_0 + \nu H^{2} + \alpha
\frac{H^{4}}{H_{I}^{2}}\right)\,,
\end{equation}
in which  $c_0$  is a constant of dimension $+2$ in natural units, closely related to $\Lambda$; $H_I$ is a dimension $+1$ scale related to inflation; and  $\nu$ and $\alpha$ are dimensionless coefficients, the former being obviously related to $\nueff$ from the previous section.  Such extended expression for the VED involving both $\sim H^2$ and $\sim H^4$ terms can  produce successful inflation with graceful exit in the early universe\,\cite{JSPRev2015,rvmInflation,GRF2015,Yu2020} and leaves an effective form of dynamical VED for the present universe behaving as (\ref{eq:RVM2}).  Remarkably, that form has been positively confronted with the data\,\cite{ApJL2015,RVMphenoOlder1,RVMpheno1,RVMpheno2,ApJL2019,Mehdi2019}.

From our direct QFT calculation, we have seen that the $\sim H^2$ terms indeed apear (see also Appendix \ref{AppendixC} for a more general case)  whereas the higher order terms that we have obtained are more along the Starobinsky inflationary line. However, we cannot exclude the presence of the  $\sim H^4$   string-induced effective contributions, as discussed in\,\cite{BMS2020A,BMS2020B}.  Being these contributions nonvanishing for $H=H_I=$const.  and taking into account that the Starobinsky-like higher order terms just vanish in such regime, it is reasonable to expect that for large values of $H_I$  the  $\sim H^4$  terms (if available from string-induced origin) prove to be the dominant terms  at the inflationary scale. If so, this could change dramatically our picture of inflation  into a more RVM-like one.

\section{Discussion and conclusions}\label{sec:conclusions}

We have devoted this paper to investigate the possible dynamics of vacuum in the context of quantum field theory in the expanding universe, and more specifically in FLRW spacetime. The quantum field theoretical context is well-known\,\cite{BirrellDavies82,ParkerToms09,Fulling89,MukhanovWinitzki07} but the difficulties are still of formidable magnitude.  This is obviously so since we know that in this kind of business sooner or later we have to  face a huge stumbling block on our way, which is the cosmological constant problem\,\cite{Weinberg89,Witten2000}. Such mystery is perhaps the greatest conceptual challenge faced by theoretical physics ever, owing to the mind boggling discrepancy existing between the measured value of the vacuum energy density (VED) and the typically predicted one by our most cherished QFT's, say quantum chromodynamics and specially the electroweak standard model, both being essential parts of  what we call the standard model of particle physics, which in itself is a highly successful theory of the fundamental interactions.  Even though tackling such problems may require the concepts and the sophisticated theoretical tools underlying quantum gravity and string theory\,\cite{Witten2000}, difficulties appear indeed in all fronts, and string theory might not be an exception. Indeed,  as of some time we known that string theories somehow abhor de Sitter space, as  `swampland' conjectures point to the impossibility to construct metastable de Sitter vacua in such theories\,\cite{Vafa}. We remain simply agnostic about these problems but, if true, they  add up more trouble to the list of conundrums that fundamental physics has to face when addressing the physics of  vacuum in an expanding universe.  In the meantime, we expect that some sort of provisional result  should perhaps be possible within the --  much more pedestrian -- semiclassical QFT approach, in which quantum matter fields interact with an external gravitational field.

Specifically, in this work  we have reconsidered the calculation of the renormalized energy-momentum tensor (EMT) of a real quantum scalar field non-minimally coupled to the FLRW background.  We have performed the calculation following two lines of approach based on adiabatic regularization and  renormalization of the EMT.  In both cases we started from the WKB expansion of the field modes in the  FLRW spacetime. Then we defined an appropriately renormalized EMT by performing a substraction  of its on-shell value (i.e. the value defined at the mass $m$ of the quantized field)  at an arbitrary renormalization point $M$. The resulting EMT becomes finite because we subtract  the first four adiabatic orders (the only ones that can be divergent).  Since the renormalized EMT  becomes a function of  the arbitrary scale $M$, we can compare the renormalized result at different epochs of the cosmic history characterized by different energy scales.   In one of the approaches (presented in the main text) we have shown by direct calculation that the renormalized EMT defined in that way is finite.  In another approach (left for Appendix \ref{AppendixB}) we use dimensional regularization to subtract the poles of the low adiabatic orders. Here we use the more conventional method based on cancelling the poles using the counterterms associated to the fundamental parameters $\rL,G_N^{-1}$ and $a_1$ (the coefficient of $R^2$).   The two approaches concur to the same renormalized result. The next important point is the extraction of the VED from the renormalized  EMT, which is composed not only of the zero-point energy part (involving the quantum fluctuations of the scalar field) but also of $\rL(M)$,  the renormalized value of $\rL$ at the scale $M$. Remarkably, the sum of these two quantities is free from quartic terms  $\sim m^4$, which are usually responsible for the exceedingly large contributions to the VED and the corresponding need for fine-tuning.

We have also shown that the renormalized VED  obtained from this QFT calculation takes on approximately the usual form of the running vacuum models (RVM's)\,\cite{JSPRev2013}, in which $\rv=\rv(H)$ appears in the manner of an additive constant plus a series of powers of $H$ (the Hubble rate) and its time derivatives.  Originally, the RVM approach  was motivated from general considerations involving the renormalization group in QFT in curved spacetime (cf. \cite{JSPRev2013} and references therein).   At the end of the day, we have been able to show that the RVM form of the VED for the current universe can be achieved from direct calculations of QFT in the FLRW spacetime.   In it, all the terms  made out of powers of $H$ (and its time derivatives)  are of even adiabatic order.  This means that all these powers  effectively carry an even number of time derivatives of the scale factor, which is essential to preserve the general covariance of the action.  The lowest order dynamical component of the VED is just  $\sim\nu\,H^2$,  where the dimensionless coefficient $\nu$ is naturally predicted to be small  ($|\nu|\ll1$), but must ultimately be determined experimentally by confronting the model to the cosmological data.  That term is nevertheless sufficient to describe the dynamics of the vacuum in the current universe, while the higher order components can play a role in the early universe, and in particular for describing inflation.
 In fact, in previous works the model has been phenomenologically fitted to a large wealth of cosmological data and the running parameter $\nu$ has been found to be positive and in the ballpark of $\sim 10^{-3}$\,\,\cite{ApJL2015,RVMphenoOlder1,RVMpheno1,RVMpheno2}.  Let us finally mention that even though our QFT calculation has been simplified by the use of a single (real) quantum scalar field, further investigations will be needed to generalize these results for  multiple fields,  involving scalar as well as  vector and fermionic components. Up to computational details, however, we expect that the main results of the renormalization program presented here should be maintained.

 \subsection*{Acknowledgements}

We are  grateful to A. G\'omez-Valent and J. de Cruz P\'erez for reading the manuscript and for useful comments.  We thank the anonymous referee for insightful remarks which have led to improve our presentation.
 We are funded by projects FPA2016-76005-C2-1-P (MINECO), 2017-SGR-929 (Generalitat de
Catalunya) and MDM-2014-0369 (ICCUB). CMP is also partially supported  by  the fellowship 2019 FI$_{-}$B 00351. The work of JSP is  also partially supported by the COST Association Action CA18108  ``Quantum Gravity Phenomenology in the Multimessenger Approach  (QG-MM)''.

\appendix
\section{Conventions and geometrical quantities}\label{AppendixA}
We use natural units, therefore $\hbar=c=1$ and $G_N=1/M_P$, where $M_P$ is the Planck mass. As for the conventions on geometrical quantities used throughout this work, they read as follows: signature of the metric  $g_{\mu\nu}$, $(-, +,+,+ )$; Riemann tensor,
$R^\lambda_{\,\,\,\,\mu \nu \sigma} = \partial_\nu \, \Gamma^\lambda_{\,\,\mu\sigma} + \Gamma^\rho_{\,\, \mu\sigma} \, \Gamma^\lambda_{\,\, \rho\nu} - (\nu \leftrightarrow \sigma)$; Ricci tensor, $R_{\mu\nu} = R^\lambda_{\,\,\,\,\mu \lambda \nu}$; and Ricci scalar,  $R = g^{\mu\nu} R_{\mu\nu}$.  Overall, these correspond to the $(+, +, +)$ conventions in the classification by Misner-Thorn-Wheeler\,\cite{MTW}.  As usual, the Einstein tensor is defined through  $G_{\mu\nu}=R_{\mu\nu}-\frac12\,R g_{\mu\nu}$ and the Einstein field equations read $G_{\mu\nu}+\CC g_{\mu\nu}=8\pi G_N\, T_{\mu\nu}$. The Christoffel symbols associated to the conformally flat metric $ds^2=a^2(\eta)\eta_{\mu\nu}dx^\mu dx^\nu$, with $\eta_{\mu\nu}={\rm diag} (-1, +1, +1, +1)$, are the following:
\begin{equation}
\Gamma_{00}^{0}=\mathcal{H},\qquad \Gamma_{ij}^0=\mathcal{H}\delta_{ij}, \qquad \Gamma_{j0}^i=\mathcal{H}\delta_j^i\,.
\end{equation}
Recalling that the relation between the Hubble rate in conformal and cosmic times is ${\cal H}=a H$, the Ricci scalar and the nonvanishing components of the curvature tensors are alternatively given by
\begin{equation}\label{eq:R}
R={6}\frac{a^{\prime\prime}}{a^3}=\frac{6}{a^2}\,(\mathcal{H}^\prime+\mathcal{H}^2)=6\,\left(\frac{\dot{a}^2}{a^2}+\frac{\ddot{a}}{a}\right)=6(2H^2+\dot{H})
\end{equation}
and
\begin{equation}\label{eq:R00G00}
 \ \ \ R_{00}=-3\mathcal{H}^\prime=-3a^2(H^2+\dot{H})\,,\qquad G_{00}=3\mathcal{H}^2=3a^2H^2\,.
\end{equation}
We remind the reader that primes indicate differentiation with respect to conformal time and dots differentiation with respect to cosmic time.
We also need the  higher order curvature tensor  (of adiabatic order $4$) obtained by functionally differentiating the $R^2$-term in the higher derivative vacuum action:
\begin{equation}\label{eq:H1munu}
  H_{\mu\nu}^{(1)}=\frac{1}{\sqrt{-g}}  \frac{\delta}{\delta g^{\mu\nu}}  \int d^4x  \sqrt{-g} R^2=-2\nabla_\mu\nabla_\nu R + 2 g_{\mu\nu} \Box R - \frac12 g_{\mu\nu} R^2 +2 R R_{\mu\nu}\,.
\end{equation}
Its  $00$-component in the conformally flat metric reads
\begin{equation}\label{eq:H100}
H_{00}^{(1)}=\frac{-18}{a^2}\left(\mathcal{H}^{\prime 2}-2\mathcal{H}^{\prime \prime}\mathcal{H}+3 \mathcal{H}^4 \right)= -18 a^2\left(\dot{H}^2-2H\ddot{H}-6H^2\dot{H}\right)\,.
\end{equation}

\section{Combining adiabatic and dimensional regularization}\label{AppendixB}
In this appendix, we  sketch the calculation of the regularized EMT by using dimensional regularization (DR).  Let us nonetheless emphasize that while we will use minimal subtraction of poles as a regularization procedure, we do \textit{not} intend to renormalize the theory with this prescription. If we would do that the renormalized vacuum energy would still exhibit the unwanted  $\sim m^4$ contributions.  In the following, we show that after the ARP has been performed,   the divergent integrals  appearing  in the intermediate calculations can be regularized through DR  and then we can recover exactly the same result \eqref{RenormalizedVE} for the renormalized VED.

\subsection{Useful formulas}
For our purposes it will suffice to  focus on integrals of the form
\begin{equation}\label{eq:MasterIntegral}
I_3(n,Q)\equiv\int \frac{d^3 k}{(2\pi)^3}\frac{1}{\omega^n_k(Q)}=\frac{1}{2\pi^2}\int dk k^2 \frac{1}{\omega^n_k(Q)}=\frac{1}{2\pi^2}\int dk k^2\frac{1}{(k^2+Q^2)^{n/2}}\,,
\end{equation}
where  $k\equiv|\bk|$ and $Q$ is an arbitrary energy scale. If we generalize it to N dimensions,
\begin{equation}
\begin{split}
I_N(n,Q)&\equiv \int \frac{d^N k}{(2\pi)^N}\frac{1}{(k^2+Q^2)^{n/2}}=\frac{1}{(4\pi)^{N/2}}\frac{\Gamma\left(\frac{n-N}{2} \right)}{\Gamma\left( \frac{n}{2}\right)}\left(\frac{1}{Q^2}\right)^{\frac{n-N}{2}}\\
&=\frac{1}{(4\pi)^{n/2}}\frac{\Gamma\left(\frac{n-N}{2} \right)}{\Gamma\left( \frac{n}{2}\right)}\left(\frac{Q^2}{4\pi}\right)^{\frac{N-n}{2}}=\frac{1}{(4\pi)^{n/2}}\frac{\Gamma\left(\frac{n-3}{2} +\epsilon\right)}{\Gamma\left( \frac{n}{2}\right)}\left(\frac{Q^2}{4\pi}\right)^{\frac{3-n}{2}}\left(\frac{Q^2}{4\pi \mu^2}\right)^{-\epsilon}. \label{DRFormula}
\end{split}
\end{equation}
where in the last step we have set  $N=3-2\epsilon$. Notice that the  scale $\mu$ has been introduced for dimensional purposes only and has no obvious physical meaning.  It could have equally well  been inserted  in the original integral in the form $d^N k\to \mu^{2\epsilon} d^N k$ to insure that the integration measure has the same dimension as $d^3k$.   Poles appear for $n\leq 3$  in the integral (\ref{eq:MasterIntegral}) and then we can use relations such as
\begin{equation}\label{eq:Gammaepsilon}
\frac{\Gamma(\epsilon)}{(4\pi)^{-\epsilon}}=\frac{1}{\epsilon} - \gamma_E +\ln (4\pi) +{\cal O}(\epsilon)\,,\ \ \ \ \  \frac{\Gamma(-1+\epsilon)}{(4\pi)^{-\epsilon}}=-\frac{1}{\epsilon} - 1+\gamma_E -\ln (4\pi) +{\cal O}(\epsilon)\,,
\end{equation}
which parameterize the divergent behavior of Euler's $\Gamma$ function near the origin and near $-1$,  respectively, where $\gamma_E$ is Euler's constant. Similar expressions can be generated to parameterize the divergent behavior of $\Gamma$ around other negative integers using the well-known functional relation  $\Gamma(x+1)= x\,\Gamma (x)$.

\subsection{Dimensionally regularized ZPE in FLRW spacetime}

Next we summarize  how to obtain the same expression for the renormalized VED as the one we have found in Sec. \ref{sec:TotalVED},  but now  using DR in the intermediate steps to regularize the divergent integrals.
Our common starting point is Eq.\, \eqref{DecompositionEMT},
\begin{equation}
\langle T_{00}^{\delta \phi}\rangle (M)= \langle T_{00}^{\delta \phi}\rangle_{Div}(M)+\langle T_{00}^{\delta \phi}\rangle_{Non-Div}(M), \label{DR_EMT}
\end{equation}
where the divergent and non-divergent contributions are the same ones as in equations\,(\ref{DivergentPart}) and (\ref{Non-DivergentPart}), respectively.  The order of adiabaticity of these expressions, therefore is the same as in the calculation presented in the main text, and we shall take this fact for granted hereafter.  We should remind the reader that in these expressions the WKB expansion of the modes has been  performed off-shell, i.e. at an arbitrary mass scale $M$ which is generally different from the physical mass, $m$.  However, at this point we take a different route for the rest of the calculation, namely  we compute the divergent parts with the help of the DR  formula \eqref{DRFormula}.  Next we expand in $\epsilon$ before taking the limit $\epsilon\to 0$ and leave only the $\epsilon$ dependence  at the poles located at $\epsilon=0$ (i.e. $N=3$).  The final result is
\begin{equation}\label{DivergentPartExplicitelyDR}
\begin{split}
\langle T_{00}^{\delta \phi}\rangle_{Div}(M)&=-\frac{M^4 a^2}{64\pi^2}\left[\frac{1}{\epsilon}+\frac{3}{2}-\gamma_E+\ln 4\pi +\ln \frac{\mu^2}{M^2}\right]\\
&-\frac{3M^2 \mathcal{H}^2}{16\pi^2}\left(\xi-\frac{1}{6} \right)\left[\frac{1}{\epsilon}-1-\gamma_E+\ln 4\pi +\ln\frac{\mu^2}{M^2}\right]\\
&-\frac{9}{16\pi^2 a^2}\left(\xi-\frac{1}{6}\right)^2(2\mathcal{H}^{\prime \prime}-\mathcal{H}^{\prime 2}-3\mathcal{H}^4)\left[\frac{1}{\epsilon}-\gamma_E+\ln 4\pi +\ln \frac{\mu^2}{M^2}\right]\\
&-\frac{\Delta^2 a^2 M^2}{32\pi^2}\left[\frac{1}{\epsilon}+1-\gamma_E+\ln 4\pi +\ln \frac{\mu^2}{M^2}\right]
-\frac{\Delta^4 a^2}{64\pi^2}\left[\frac{1}{\epsilon}-\gamma_E+\ln 4\pi +\ln \frac{\mu^2}{M^2}\right]\\
&-\left(\xi-\frac{1}{6}\right)\frac{3\Delta^2 \mathcal{H}^2}{16\pi^2}\left[\frac{1}{\epsilon}-\gamma_E+\ln 4\pi +\ln \frac{\mu^2}{M^2}\right]
\end{split}
\end{equation}
This equation can be conveniently split into a UV-divergent part involving the poles at $\epsilon=0$ and a finite part. Defining
\begin{equation}\label{eq:Depsilon}
  D_\epsilon=\frac{1}{\epsilon}-\gamma_E+\ln 4\pi
\end{equation}
and recalling that $\Delta^2=m^2-M^2$, we obtain
\begin{equation}\label{eq:split1}
\begin{split}
\langle T_{00}^{\delta \phi}\rangle_{Div}(M)&=-\frac{m^4 a^2}{64\pi^2}D_\epsilon-\frac{3 m^2\cH^2}{16\pi^2} \left(\xi-
\frac16\right) D_\epsilon
-\frac{9}{16\pi^2 a^2}\left(\xi-\frac16\right)^2 (2\mathcal{H}^{\prime \prime}-\mathcal{H}^{\prime 2}-3\mathcal{H}^4)D_\epsilon\\
&+\langle T_{00}^{\delta \phi}\rangle_{\rm FR}(M)\,.
\end{split}
\end{equation}
The UV-divergent part, in the first line, depends only on the physical mass of the particle, $m$, whereas the finite remainder (denoted with the label FR) depends  both on the mass and on the renormalization point $M$:
\begin{eqnarray}
\langle T_{00}^{\delta \phi}\rangle_{\rm FR}(M)&=&-\frac{M^4 a^2}{64\pi^2}\left[\frac{3}{2}+\ln\frac{\mu^2}{M^2}\right]-\frac{3M^2 \mathcal{H}^2}{16\pi^2}\left(\xi-\frac{1}{6}\right)\left[ -1+\ln\frac{\mu^2}{M^2} \right]\nonumber\\
&&-\frac{\Delta^2 a^2 M^2}{32\pi^2}\left[1+\ln \frac{\mu^2}{M^2}\right]-\frac{\Delta^4 a^2 }{64\pi^2}\ln \frac{\mu^2}{M^2}
-\left(\xi-\frac{1}{6}\right)\frac{3\Delta^2 \mathcal{H}^2}{16\pi^2}\ln \frac{\mu^2}{M^2}\nonumber\\
&&-\frac{9}{16\pi^2 a^2}\left(\xi-\frac{1}{6} \right)^2(2\mathcal{H}^{\prime \prime}\mathcal{H}-\mathcal{H}^{\prime 2}-3\mathcal{H}^4)\ln\frac{\mu^2}{M^2}\label{FRi}\\
&=&\frac{a^2}{128\pi^2}\left(M^4-4m^2M^2-2m^4\ln \frac{\mu^{2}}{M^2}\right) \nonumber\\
&&+\frac{3}{16\pi^2}\left(\xi-\frac{1}{6}\right) \cH^2\left(M^2 -m^2\ln \frac{\mu^{2}}{M^2}\right)\nonumber\\
&&-\frac{9}{16\pi^2 a^2}\left(\xi-\frac{1}{6} \right)^2(2\mathcal{H}^{\prime \prime}\mathcal{H}-\mathcal{H}^{\prime 2}-3\mathcal{H}^4)\ln\frac{\mu^2}{M^2}\,,\label{FRii}
\end{eqnarray}
where in the second equality we have used once more $\Delta^2=m^2-M^2$. At this stage,  the DR procedure carries a dependence on the artificial mass scale $\mu$.  However, in our case $\mu$ will play no role since we are not just aiming at a conventional renormalization based on minimal subtraction, so $\mu$ serves only as an auxiliary variable which will eventually disappear from the renormalized result.  We should emphasize that the  relevant renormalization scale in our calculation is not $\mu$ but  $M$.   Dimensional regularization is used here only as a technique to display explicitly the divergences of the EMT  and to enable their subtraction with the conventional counterterm procedure.

\subsection{Counterterms}

While the calculation can be  fully carried out without any use of DR,  provided one defines a properly subtracted EMT from the beginning with the ARP (cf. Sec. \ref{sec:RenormZPE}), we follow now the more conventional approach. Thus  we remove the unphysical divergences of the EMT by generating  counterterms from the coupling constants present in the extended gravitational action with the HD terms. The modified Einstein's equations read formally as in Eq.\,(\ref{eq:MEEs}) but carrying  the bare couplings, i.e. couplings which are formally UV-divergent and scale independent:
\begin{equation}
\frac{1}{8\pi G_N}G_{\mu \nu}+\rho_\Lambda g_{\mu \nu}+a_1 H_{\mu \nu}^{(1)}=\langle T_{\mu \nu}^{\delta \phi} \rangle +T_{\mu \nu}^{\phi_b}\,.
\end{equation}
We will focus on the $00$-component of this equation since we are interested in the ZPE.

Following the standard  renormalization procedure, we split each of the bare couplings on the \textit{l.h.s} of the above equation  into the renormalized term (which depends on the renormalization point $M$), and a counterterm (which does not depend on $M$):
\begin{equation}\label{eq:splitcounters}
\begin{split}
G_N^{-1}&= G_N^{-1}(M)+\delta G_N^{-1},\\
\rho_\Lambda&=\rho_\Lambda(M)+\delta\rho_\Lambda,\\
a_1&=a_1(M)+\delta a_1.
\end{split}
\end{equation}
We define the counterterms such that we can subtract the universal terms $\gamma_E$ and $4\pi$ of the DR procedure alongside with the poles, as it is conventional in the modified MS (or $\overline{\rm MS}$)\,\cite{Donoghue92,Manohar97}.  That is why we have defined the quantity $D_\epsilon$ in Eq.\,\eqref{eq:Depsilon}.
As we can see, three `primitive divergences' appear in the unrenormalized form of the EMT,  which are  proportional  to $\sim m^4$, $\sim m^2 (\xi-1/6)$  and  $(\xi-1/6)^2$, respectively. These  can be  cancelled by the corresponding counterterms  generated from the bare couplings in Eq.\,\eqref{eq:splitcounters}, i.e. the counterterms  can now be precisely used to cancel the three divergent quantities proportional to $D_\epsilon$ in Eq.\,\eqref{eq:split1}.   Using the $00$-components of the geometric tensors given in Appendix \ref{AppendixA}, they are readily found to be
\begin{equation}\label{eq:counters}
\begin{split}
  \delta G_N^{-1} &=-\frac{m^2}{2\pi}\,\left(\xi-\frac{1}{6}\right)\, D_\epsilon\,,\\
  \delta\rho_\Lambda &=+\frac{m^4}{64\pi^2}\, D_\epsilon\,,\\
  \delta a_1 &= -\frac{1}{32\pi^2}\,\left(\xi-\frac{1}{6}\right)^2\, D_\epsilon\,.
  \end{split}
\end{equation}
We confirm that  they depend on the physical mass $m$ and not on the renormalization point  $M$.
The renormalized Einstein equation resulting from cancelling the poles with the counterterms take on the same form as in Eq.\,(\ref{eq:MEEs}), in which the couplings are now the renormalized ones and explicitly depend on the mass scale $M$. The $00$-component reads
\begin{equation}\label{Einsteinequationmuind}
\frac{1}{8\pi G_N (M) } G_{00}+\rho_\Lambda (M) g_{00}+a_1 (M) H_{00}^{(1)}= \langle \widetilde{{T}_{00}^{\delta \phi}}\rangle (M)+T_{00}^{\phi_b},
\end{equation}
where the tilded quantity
\begin{equation}\label{eq:T00RenDR}
  \langle \widetilde{{T}_{00}^{\delta \phi}}\rangle (M)= \langle T_{00}^{\delta \phi}\rangle_{\rm FR}(M)+\langle T_{00}^{\delta \phi}\rangle_{Non-Div}(M)\,.
\end{equation}
 is  the finite part  left of the EMT \eqref{DR_EMT}  after removing the poles. Being finite we might be tempted to call it  (provisionally) the renormalized ZPE, but in fact  is not our final renormalized expression.   Some further  insight on it can be achieved by considering the term labelled  FR, which is  given by  (\ref{FRi}).
If we apply the limit $a=1$ (so that $\cH$ and all its derivatives vanish) and project the result on-shell ($M=m$, hence $\Delta=0$), the whole expression \eqref{FRi} or \eqref{FRii} shrinks to just one of the equivalent forms
\begin{equation}\label{eq:ZPEflat}
 \left. \langle T_{00}^{\delta \phi}\rangle_{\rm FR}(m)\right|_{\rm Minkowski}=-\frac{m^4}{64\pi^2}\left[\frac{3}{2}+\ln\frac{\mu^2}{m^2}\right]=\frac{m^4}{128\pi^2}\left[-3-2\ln\frac{\mu^2}{m^2}\right]=\frac{m^4}{64\pi^2}\left[\ln\frac{m^2}{\mu^2}-\frac{3}{2}\right]\,.\phantom{XX}\
\end{equation}
This  is nothing but the standard  (one-loop) ZPE in flat spacetime, namely it is the renormalized form of the UV-divergent integral (\ref{eq:Minkoski}) within the  $\overline{\rm MS}$.  As we can see, Eq.\,\eqref{eq:ZPEflat} brings a explicit dependence on $\mu$ and above all  it grows as the quartic power of  the mass of the field.  Because the total VED  is the sum of  \eqref{eq:ZPEflat} plus the renormalized $\rL$  --  cf. Eq.\,(\ref{EMTvacuum}) -- we are led to face a  huge contribution from the  quartic term $\sim m^4$  (for virtually every known particle, except a very light neutrino), which  amounts to a large fine-tuning between these two quantities.  This is odd, in fact unacceptable.  As discussed in detail in \cite{JSPRev2013}, the flat space formula carries indeed the core of the cosmological constant problem\,\cite{Weinberg89} and the curved spacetime calculation just inherits it at this point, but it does not aggravate it further. Thus, not surprisingly the subtraction of this part leaves a well-behaved result (cf. Secc. \ref{Sec:SubtractMinkowski}).  However, let us continue with our renormalization procedure and evade this conundrum within the present context.

\subsection{Renormalized ZPE and absence of $\sim m^4$ contributions}

The problem stems from the tilded definition of the renormalized EMT given in \eqref{eq:T00RenDR},  which is just a variant of the  $\overline{\rm  MS}$-renormalized one, although carrying off-shell $\Delta^2$-corrections.  However, a well-defined expression can be obtained if we call back anew our definition of renormalized EMT as in Eq.\,(\ref{EMTRenormalized}) of the main text.  The prescription amounts to take the on-shell value (at the physical mass $m$) and subtract from it the terms up to 4th adiabatic order  at some arbitrary mass scale $M$.  This provides automatically  an overall finite result, as we have proven in the main text without using DR. Taking  into account that in this alternative procedure we have already removed the poles appearing in the intermediate steps with the help of DR,  it suffices to perform the aforementioned subtraction directly with the  finite expression \eqref{eq:T00RenDR}:
\begin{eqnarray}\label{EMTRenormalized2}
\langle T_{00}^{\delta \phi}\rangle_{\rm Ren}(M)&=& \langle \widetilde{{T}_{00}^{\delta \phi}}\rangle(m) -  \langle \widetilde{{T}_{00}^{\delta \phi}}\rangle (M)\nonumber\\
&=&\langle T_{00}^{\delta \phi}\rangle_{\rm FR}(m)-\langle T_{00}^{\delta \phi}\rangle_{\rm FR}(M)+\langle T_{00}^{\delta \phi}\rangle_{\rm Non-div}(m)-\langle T_{00}^{\delta \phi}\rangle_{\rm Non-div}(M)\nonumber\\
&=&\langle T_{00}^{\delta \phi}\rangle_{\rm FR}(m)-\langle T_{00}^{\delta \phi}\rangle_{\rm FR}(M)-\left(\xi-\frac{1}{6}\right)\frac{3\Delta^2 \cH^2}{8\pi^2}+\dots
\end{eqnarray}
Upon some simple rearrangements, it finally yields
\begin{eqnarray}\label{eq:RenormalizedFinal}
&&\langle T_{00}^{\delta \phi}\rangle_{\rm Ren}(M)=\frac{a^2}{128\pi^2 }\left(-M^4+4m^2M^2-3m^4+2m^4 \ln \frac{m^2}{M^2}\right)\nonumber\\
&&-\left(\xi-\frac{1}{6}\right)\frac{3 \mathcal{H}^2 }{16 \pi^2 }\left(m^2-M^2-m^2\ln \frac{m^2}{M^2} \right)
+\left(\xi-\frac{1}{6}\right)^2 \frac{9\left(2  \mathcal{H}^{\prime \prime} \mathcal{H}- \mathcal{H}^{\prime 2}- 3  \mathcal{H}^{4}\right)}{16\pi^2 a^2}\ln \frac{m^2}{M^2}+\dots\nonumber\\
\end{eqnarray}
The $\mu$-dependence has cancelled at this point, and as we can see this equation turns out to be exactly the same one as in Eq.\eqref{RenormalizedExplicit2}. Therefore, from this point onwards  we can reproduce the same renormalized VED  \eqref{RenormalizedVE}, just starting from \eqref{eq:Totalrhovac} and subtracting its value at the two scales  $M$ and $M_0$.  Once more the result is that the VED at the scale $M$ can be related with its value at another scale $M_0$ without receiving any contribution from the quartic values of the mass scales or  of the mass of the particle.  Thus, on using this renormalization procedure we can get rid of  the dependence on the quartic powers of the masses as well as on the spurious  DR parameter $\mu$.

The lesson we can learn is the following.  While the mere $\overline{\rm  MS}$ renormalization of the  VED (based on using DR together with the subtraction of the poles by the counterterms)  leaves  a result which is explicitly  dependent both on the artifical DR scale $\mu$ and on the quartic powers of the masses\,\cite{KohriMatsui2017}, the extended ARP technique\,\cite{Ferreiro2019} allows to relate the renormalized quantities at different scales. With detailed calculations, which we have presented here through two different approaches (one of them not using DR at all), we have shown that  we  can avert  the mentioned problems associated to a mere removal of the poles by the counterterms.  The common final result emerging from the two procedures is an expression for the  running of the renormalized EMT in  a FLRW background as a function of the Hubble rate,  thus allowing to trace the VED evolution throughout the cosmic history.  The result we have obtained is indeed much closer in spirit to the renormalization group approach of the RVM,  cf.\,\cite{JSPRev2013,JSPRev2015} and references therein -- particularly \cite{ShapSol1,Fossil2008,ShapSol2} --  in which such mild evolution of the vacuum energy density in terms of (even) powers of the Hubble rate was predicted on very general grounds. Here we have provided for the first time a detailed account from explicit QFT calculations under an appropriate renormalization scheme leading to a possible physical interpretation of the results.  The outcome is Eq.\,\eqref{RenormalizedVE}.

We should perhaps repeat once more that such relation is \textit{not} a prediction of the value of the CC and in general of the VED,  as this is out of the scope of renormalization theory.  Every renormalization calculation needs a set of renormalization conditions. Behind these renormalization conditions there is a set of physical (and hopefully known)  inputs and from these observational inputs we can predict other physical results.  In the present instance, this means that given the VED at one scale (entailing a physical input) we can predict its value at another scale. What is, however, distinctive in the kind of calculation we have presented here is the fact that the connection between the renormalized values of the VED at different points appears smooth enough, i.e. it does not involve $\sim m^4$ terms, which are usually very large for ordinary particle masses in the standard model of particle physics (let alone in GUT's) and this suggests that no fine tuning is actually involved.

The unsatisfactory status of the $m^4$ terms in cosmology  is very similar to the hierarchy problem associated to the $m^2$ terms in ordinary gauge theories\,\cite{Veltman1981}, but even worse in magnitude.  In stark contrast to the usual situation with these terms, in the approach we have outlined in this work we do not need to call for special cancellations  (fine-tuning) among the various $m^4$ contributions from different particles, such as e.g. when using the Pauli sum rules -- see  \,\cite{Visser2018} for a detailed discussion --, nor to invoke the existence of emergent scales or very small dimensionless parameters suppressing the undesired effects, see e.g.\,\cite{Bjorken2011,Jegerlehner2014,Elahe2016,Bass2020} for a variety of contexts of this sort.  The problem is fixed here automatically by the renormalization process itself that we have used.

\section{Identification of the vacuum energy density.}\label{AppendixC}

In Minkowskian spacetime we know that the EMT  of vacuum takes on the form  $ T_{\mu \nu}^{\rm vac} =-\rho_{\rm vac} \eta_{\mu\nu}$, being the Lorentz metric $\eta_{\mu\nu}$ the only geometric structure available in a flat background to construct a Lorentz-invariant quantity which can characterize the vacuum state. This allows us to identify the vacuum energy density (VED) from the general structure of the vacuum EMT.   It is natural to generalize this identification by assuming that in curved spacetime the vacuum EMT should take the form  $ T_{\mu \nu}^{\rm vac} =-\rho_{\rm vac} g_{\mu\nu}$, with $g_{\mu\nu}$ the general background metric. We can formally motivate this result by starting from the EH action
\begin{equation}\label{eq:EH}
S_{\rm EH}=  \frac{1}{16\pi G_N}\int d^4 x \sqrt{-g}\, R  -  \int d^4 x \sqrt{-g}\, \rho_{\rm vac}\,.
\end{equation}
Varying the part involving  the vacuum energy density (i.e. the second term on the $\textit{r.h.s}$, which we call $ S_{\rm vac}$) yields
\begin{equation}\label{eq:VacuumEH}
\delta S_{\rm vac}= -\int d^4 x \,\delta \sqrt{-g}\, \rho_{\rm vac} =  -\frac{1}{2}\int  d^4 x\, \sqrt{-g}\, \left(- \rho_{\rm vac}\, g_{\mu\nu}\right)\,\delta g^{\mu\nu}\equiv -\frac{1}{2}\int  d^4 x\, \sqrt{-g}\, T_{\mu \nu}^{\rm vac} \delta g^{\mu\nu} \,,
\end{equation}
which provides the identification $ T_{\mu \nu}^{\rm vac} =-\rho_{\rm vac} g_{\mu\nu}$.
This is the line of approach that we have followed here.  However, we should point out that such identification has some ambiguities, which as we shall argue below should not alter in a significant way the results that we have obtained and,  remarkably enough,   lead to a generalized form of the RVM which had actually been considered previously in the literature in different phenomenological formulations\,\cite{ApJL2015,RVMphenoOlder1}. In this sense we believe this point deserves being mentioned  here,  see also\,\cite{Maggiore2011,Bilic2011, Bilic2012}.

\subsection{More geometric structures for vacuum in curved spacetime}

As we know, the vacuum effective action of QFT in the presence of gravity contains the higher derivative terms\,\cite{BirrellDavies82,ParkerToms09}.
This is because in  curved spacetime we have more geometric quantities to characterize the vacuum, and these structures  are actually necessary for implementing the renormalizability  of the semiclassical theory of quantized matter fields in an external gravitational background, as we have just seen in our discussion in Appendix \ref{AppendixB}. By the same token one might expect a more general relation between the  VED and  the EMT,  which we may schematize as follows:
\begin{equation}\label{eq:GeneralVDE1}
  T_{\mu \nu}^{\rm vac}=-\rho_{vac} g_{\mu\nu}-\alpha_1R  g_{\mu\nu}-\alpha_2R_{\mu\nu}+ {\cal O}(R^2)\,,
\end{equation}
in which ${\cal O}(R^2)$ stand for the higher derivative terms, and $\alpha_i$ are parameters of dimension $+2$ in natural units. For $g_{\mu\nu}=\eta_{\mu\nu}$  the previous ansatz just boils down to the flat spacetime form mentioned above. To illustrate the possible impact of the additional terms, let us first focus on the following specific form for the renormalized energy-momentum tensor:
\begin{equation}\label{eq:GeneralVDE2}
  T_{\mu \nu}^{\rm vac}=-\rho_{vac} g_{\mu\nu}+\frac{\lambda}{16\pi^2} M^2 G_{\mu\nu}+ {\cal O}(R^2)\,.
\end{equation}
Here,  $G_{\mu\nu}$ is Einstein's tensor (cf. Appendix \ref{AppendixA}).  We restrict hereafter our considerations to the late universe since we wish to assess what is the possible impact of the new terms on the parameters that can be directly fitted to observations. The appearance of the mass scale $M$ is necessary for dimensional reasons. Furthermore,  $\lambda$ is an appropriately normalized (dimensionless) parameter. 
On equating this expression to Eq.\,(\ref{EMTvacuum}) and considering the $00$-component, we find a generalized form of (\ref{eq:Totalrhovac}):
\begin{equation}\label{eq:Totalrhovac2}
\rho_{vac}(M)=\rho_{\Lambda}(M)+\frac{\langle T_{00}^{\delta \phi}\rangle_{\rm Ren}(M )}{a^2}-\frac{3\lambda}{16\pi^2} M^2 H^2\,.
\end{equation}
If we repeat the same steps that led us to  Eq.\,\eqref{eq:RVM2}, but keeping the additional term in (\ref{eq:Totalrhovac2}), we arrive at a very similar result  (\ref{eq:nueff})  for the VED, except that the effective running parameter receives also a contribution from $\lambda$. Specifically, we find
\begin{equation}\label{eq:nueff2}
\nueff=\frac{1}{2\pi}\,\frac{M_X^2}{M_P^2}\,\left[\left(\frac{1}{6}-\xi\right)\left(1+\frac{m^2}{M_X^2}\ln \frac{H_0^{2}}{M_X^2}\right)+\lambda\right]\,.
\end{equation}

Thus, formally the expression for the VED is the same, except that the new term from (\ref{eq:Totalrhovac2}) resulted in a new contribution to the effective coefficient $\nueff$, which in any case must be fitted to the experiment irrespective of its inner theoretical structure built up from different sectors.

\subsection{Generalized form of the RVM}

Let us finally consider the modification introduced by the more general form \eqref{eq:GeneralVDE1}.  For convenience we first redefine the (dimension +2)  coefficients of that expression as $\alpha_i=\frac{\lambda_i}{16\pi^2} M^2\, (i=1,2)$.  Using the explicit form for $R$ and  $R_{00}$ in the conformally flat metric as given in Appendix \ref{AppendixA}, we obtain after a straightforward calculation:
\begin{equation}\label{eq:RVMgeneralized}
\rho_{vac}(H)=\rho_{vac}(M_X)+\frac{3}{8\pi G_N}\,\left(\tilde{\nu}_{\rm eff}(H)\,H^2+ \bar{\nu}\,\dot{H}\right)\,,
\end{equation}
where
\begin{equation}\label{eq:nuefftilde}
\tilde{\nu}_{\rm eff}(H)=\frac{1}{2\pi}\,\frac{M_X^2}{M_P^2}\,\left[\left(\frac{1}{6}-\xi\right)\left(1+\frac{m^2}{M_X^2}\ln \frac{H^{2}}{M_X^2}\right)+4\lambda_1+\lambda_2\right]\,.
\end{equation}
and
\begin{equation}\label{eq:nubar}
\bar{\nu}=\frac{2\lambda_1+\lambda_2}{2\pi}\,\frac{M_X^2}{M_P^2}\,.
\end{equation}
We may neglect as before the log evolution of $\tilde{\nu}_{\rm eff}(H)$ and approximate   $\tilde{\nu}_{\rm eff}(H_0)\simeq \tilde{\nu}_{\rm eff}$.  The coefficient $\bar{\nu}$ is dealt with as constant here, but in general the dimensionless couplings $\lambda_i$ may  also be dependent on the scale $M$, although we expect that the renormalization effects should be logarithmic; and,  therefore, following the same practice as with $\tilde{\nu}_{\rm eff}(H)$,   we have not considered these subleadig effects here for the sake of a simpler presentation.
We can easily verify that for $\lambda_1=\lambda/2$ and $\lambda_2=-\lambda$ the last two formulas reduce to \eqref{eq:nueff2} and  $\bar{\nu}=0$, respectively,   since in that case \eqref{eq:GeneralVDE1} boils down to \eqref{eq:GeneralVDE2}.
Finally,  determining $\rho_{vac}(M_X)$ from the boundary condition $\rho_{vac}(H_0)=\rvo$, we can write down \eqref{eq:RVMgeneralized} in a manifestly normalized way with respect to the current values:
\begin{equation}\label{eq:RVMgeneralized2}
\rho_{vac}(H)\simeq\rvo+\frac{3\tilde{\nu}_{\rm eff}}{8\pi G_N}\,(H^2-H_0^2)+ \frac{3\bar{\nu}}{8\pi G_N}\,(\dot{H}-\dot{H}_0)\,,
\end{equation}
in which $H_0$ and  $\dot{H}_0$ stand, of course, for the respective values of  $H$ and $\dot{H}$ at present.  The above formula generalizes Eq.\,\eqref{eq:RVM2} by including the additional coefficient $\bar{\nu}$, which accompanies the new dynamical term $\sim\dot{H}$. We note that several generalized  forms of the  RVM  containing  dynamical terms beyond the canonical one $H^2 $ were studied under different phenomenological conditions, and fitted as well  to the data, in \cite{ApJL2015,RVMphenoOlder1}. Here we have shown that these extended forms of the RVM, which were confronted to observations in the aforementioned references, can also appear as a result of  the QFT calculations in the FLRW background.


\end{document}